\documentclass[%
 reprint,
showpacs,
 amsmath,amssymb,
 aps,
]{revtex4-1}

\pdfoutput=1

\usepackage{graphicx}
\usepackage{dcolumn}
\usepackage{bm}
\usepackage{xcolor}

\begin{document}

\preprint{APS/123-QED}

\title{Phase-space dynamics of minimal Canham-Helfrich cells}

\author{Ana M. Maitin$^{1,2}$}
\author{Francisco Monroy$^{1,2}$}%
 \email{Corresponding author: monroy@ucm.es}
\affiliation{%
1 Department of Physical Chemistry I, Complutense University, Av. Complutense s/n, 28040 Madrid, Spain \\
2 Unit of Translational Biophysics, Institute of Biomedical Research Hospital Doce de Octubre, Av. Andalucia s/n, 28041 Madrid, Spain\\
}%


\date{\today}

\begin{abstract}
The dynamical phase-space of axisymmetric Canham-Helfrich (CH) cells is constructed from a Hamiltonian field recapitulating membrane curvature-elasticity and systemic restrictions. Guiding principles are reparametrization to convert a static geometric system into a dynamical system, and Galilean transformation, to build a transformed Lagrangian invariant with respect to the action described by the CH free-energy. Building on the fluidity postulate, this Lagrangian describes the cellular membrane as an inverted harmonic oscillator driven by bending elasticity and effective friction governed by Gaussian curvature. To close the spring-mass interaction, we explicit the mass of the membrane and establish a dimensionally-minimal Lagrangian. Then, the canonical Hamiltonian is constructed in generalized coordinates $H(p, q, t)$, and the equations of motion derived in accordance with the principle of minimal action. The derived phase-space is used as a global predictor of the cellular shapes for different mechanical settings with a biological significance.

\end{abstract}

\maketitle 


Cell mechanics is a key regulator of biological function \cite{1,2,3}, playing crucial roles in cellular shape remodeling \cite{4} and tissue biomechanics \cite{5}. Furthermore, cells make functional use of mechanical signals to communicate each other \cite{6,7}, and different cellular mechanical cues are crucially involved in a range of organic processes spanning from embryogenesis \cite{4,8}, wound healing \cite{9}, cancer cell pathogenesis \cite{10} and development \cite{11}. Similar mechanical principles are likely confronted in cellular aging \cite{12} and regenerative medicine \cite{13}. Cell mechanics has entered indeed an exciting era of translational investigation in which analytic mechanics might play a key-enabling role not only for the best understanding of the underlying physical mechanisms involved in cell functioning \cite{14}, but also in the development of predictional models of mechanical cell behavior \cite{15,16}. However, thriving on those many applications involves a comprehensive physical understanding that is still far from being achieved. Facing such a challenge requires going behind the molecular complexity of cell biomechanics \cite{13,17}, to make emerging the basic material ingredients and the fundamental interactions involved in the functional dynamics of the cell \cite{18,19,20}, which may be then consistently studied within an adequate analytical framework \cite{21,22,23}. 

Analytic cell mechanics grounds on the classical Canham-Helfrich (CH) theory, which stems from an energy-density functional that describes the cellular membrane as a continuous elasticity field defined in terms of curvatures \cite{24,25}. This classical field captures the physically-relevant material properties of real biomembranes; specifically, flexural elasticity, in-plane incompressibility and lateral fluidity \cite{26,27}. The main strength of the CH theory stands on its predictive capacity of the preferred cellular shapes \cite{22,28,29,30,31}. Also other theoretical extensions have been derived from the CH cell model, including the mechanics of shape transformations \cite{32,33,34,35}, stability analyses \cite{36,37}, and dynamical fluctuations \cite{38,39,40}. Furthermore, formal CH-approaches to an analytic theory of membrane dynamics have been also pursued \cite{22}, from the Lagrangian formulation \cite{32}, through detailed geometrical analyses of the curvature-elasticity field \cite{41,42} and field-theory formulations \cite{43,44}, to explicit dynamical developments framed in the Hamilton's formalism \cite{45,46}. Despite those efforts, a canonical formulation for the Hamiltonian dynamics of CH-cells has been not delivered yet in operational terms. 

Here, we adopt a genuine mechanical standpoint, and building upon the Lagrangian formulation of the CH-theory \cite{31,32}, a canonical Hamiltonian is constructed compatible with the stiffness potential imposed by the curvature-elasticity field. Then, the phase-space dynamics is obtained as pathways compatible with the Hamilton principle of minimal action, a picture equivalent to describe the evolution of the cell profiles towards the equilibrium status. This dynamics is subjected to systemic restrictions; specifically, spontaneous curvature ($c_0$), membrane tension ($\Sigma$) and cell pressure ($P$). The phase-space perspective allows a mapping of cell energetics, which delivers a global depiction of the cellular dynamics. This is advantageous with respect to conventional approaches in terms of equilibrium shapes at static configurations from which the time-domain is completely inaccesible. Our objective consists to describe the dynamics of model CH-cells from a dimensionally reduced Hamiltonian defined in canonical coordinates (Section IV). This canonical Hamiltonian will be derived from a generating Lagrangian (Section III) constructed upon addition kinetic terms to the CH-free-energy, which is reinterpreted as the action of the cellular field of curvature-elasticity (Section II). The phase space dynamics is mapped in different material settings (Section V), and the cell shapes obtained as the integral of the phase-space trajectory corresponding to determined initial conditions and given constitutive parameters (Section VI). To the best of our knowledge, this is the first time that the phase-space of model cells is explicitly explored, an achievement that opens the gate to new opportunities of predictional analysis in cell dynamics. 

\section{\label{sec.1}The model: Canham-Helfrich cell}

In a minimal depiction, a model cell can be reduced to a shape-deformable fluid body of volume $V$ enclosed by a flexible membrane of surface area $A$. Within this description, Canham \cite{24} and Helfrich \cite{25} recognized the leading role of the curvature-elasticity field in determining the energy of the membrane, which is described as a two-dimensional sheet ($\Omega$) embedded in a 3D-Euclidean space \cite{47}. Differential geometry appears to be the adequate tool to represent the field in $\Omega$, which is assumed to constitute a differentiable manifold, with metrics $\boldsymbol{g}$, where the physical history of the system can be traced as pathways in space and time \cite{48}. Within this description, the kinematic properties of the membrane are determined by the geometrical degrees of freedom of the representing surface, specifically, its local intrinsic curvatures \cite{47}. Furthermore, the surface is considered laterally fluid \cite{26}, a determinant property of biological membranes that allows their mechanical description as material points moving embedded in an homogenous two-dimensional space \cite{49}. The fluidity property appears to be a crucial postulate; its natural consequence is the invariance of the energy under reparametrizations of the representing surface \cite{49,50}. Additionally, cell deformations are subjected to geometrical constraints; for instance, $A$ and $V$ are considered constant under isothermal conditions \cite{22}. 

\textit{Free energy.} To calculate the elastic energy of the membrane corresponding to a given cell shape, the CH-model assumes a linear curvature-elasticity field (harmonic-like) \cite{25,26}, which is consistently defined in terms of the two geometric invariants of the curvature tensor $\boldsymbol{h}$ defined for $\Omega$; these are, the mean curvature, given by the trace, $H=g^{ij} h_{ij}/2$, and the Gaussian curvature, given by the determinant $K=\det h_{ij}/\det g_{ij}$ \cite{47}. In order to explicit the total amount of energy that is convertible into mechanical work, the free energy of the whole cell is calculated as the global elastic energy of the enclosing membrane. This is subjected to physical constraints of constant $A$ (maintained under lateral tension $\Sigma$), and constant $V$ (for the fluid enclosed at a pressure difference $P$ between the outer milieu and the cell interior). Under these conditions, the free-energy functional for the CH-cell reads as \cite{28}:
\begin{equation}
\label{eq.1}
\begin{aligned}
F_{CH}(\kappa,\kappa_G,c_0;A,V)=&\int_{\Omega} dS\ \left[ \frac{\kappa}{2} \left( H-c_0 \right)^2+ \kappa_G K \right] \\
& + \Sigma A + PV,
\end{aligned}
\end{equation}
which is defined for the manifold $\Omega$, geometrically representing the cellular membrane; $dS$ is the surface element. 
 \begin{figure}
 \centering
 \includegraphics[width=9cm,height=10cm]{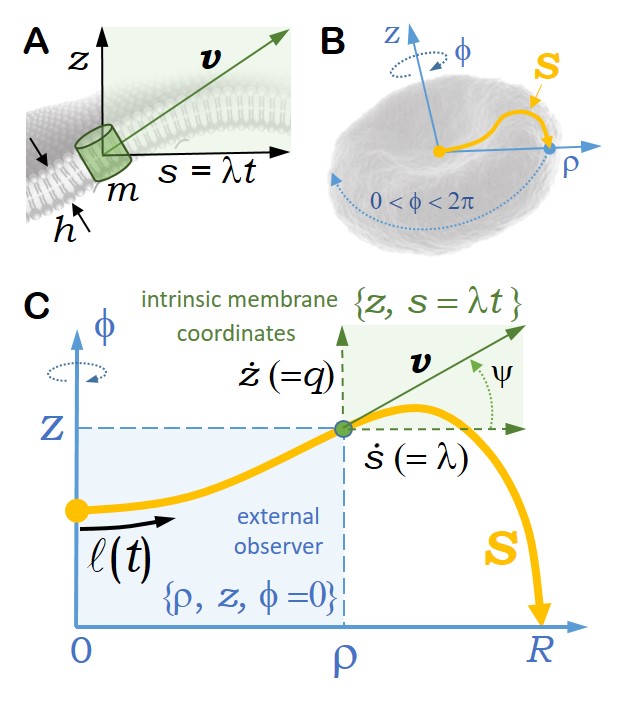}
 \caption[Texto para la lista]{$A$) Surface coordinates that define the kinematics of a generic membrane particle moving in the manifold $\Omega$. Such membrane particle is assumed with a mass $m$ and the physical dimensions of a membrane element, with a size compatible with the membrane thickness $h$. $B$) The cell surface has axial symmetry defined in a cylindrical frame of reference $\{ \rho, z,\phi \}$. A meridian pathway $\boldsymbol{S}(\rho, z)$ (in yellow) is obtained with the trajectory of this particle, travel in the parameter of evolution $ \ell $; all the meridians are assumed to be kinematically equivalent (axial symmetry), so the cell geometry is obtained by rotating around the $z$-axis. $C$) Resulting curve in cylindrical coordinates $ \{ \rho, z, \phi = 0 \} $ with $\rho \in (0, R]$ where $R$ is the equatorial cell radius.}
 \label{fig.1}
 \end{figure} 

\textit{Constitutive properties.} There are two different flexural moduli associated with the local curvatures, the bending rigidity $\kappa$, which accounts for the amount of energy necessary to bend the surface, and the Gaussian modulus $\kappa_G$, which stands for the energy associated to saddle points, where the local Gaussian curvature becomes locally negative. Both elastic moduli share a common constitutive parent, this is molecular cohesion \cite{22,51}. Whereas the bending rigidity is a positive defined quantity $(\kappa > 0)$, the Gaussian modulus usually take negative values: for lipid bilayers $\kappa_G \approx -\kappa$ \cite{52}, and $\kappa_G \approx 0$ for bicontinuous phases \cite{39}. Arguably, $\kappa_G > 0$ corresponds to membranes for which splay-saddle modes take a stabilising role. The energy scale of the elasticity field is determined by $\kappa$, which determines the energy necessary to bend the membrane. This is qualified by $\kappa_G$ describing the work involved in creating saddle regions. Furthermore, the Gauss-Bonnet theorem declarates the Gaussian curvature $K$ to be a topological invariant \cite{47}, {\it i.e.} its integrated value in a closed surface $\Omega$ remains constant at $\int_{\Omega} dS\ K =2 \pi \chi (\Omega)$, with $\chi$ being the Euler characteristic, a number that describes $\Omega$'s topology (number of holes) regardless how the surface is bent \cite{47,54}. Consequently, if the surface is deformed, its Euler characteristic ($\chi (\Omega) =2$ for the sphere) will not change \cite{42, 46}. Although the global contribution from the Gaussian curvature is conserved, any description of the local dynamics requires its  contribution to be explicitly considered. The spontaneous curvature ($c_0$) describes the natural tendency of the membrane to bend as a result of possible material asymmetries between its two faces \cite{53}. Transversely symmetric sheets stand at the flat configuration, {\it i.e.} $c_0 = 0$, whereas asymmetric membranes are characterized, either by $c_0 > 0$ if are prone to take a convex curvature (with respect to the cell interior), or by $c_0 < 0$ if concave \cite{22,53}. The systemic parameters $\Sigma$ and $P$ are Lagrange multipliers (as defined in Eq. (1)), which impose the aforementioned area and volume constraints. The vacuum state represents a tensionless ($\Sigma = 0$) and floppy ($P = 0$) membrane with its mean curvature in equilibrium at $c_0$, with an energy at the absolute minimum of Eq. \eqref{eq.1}. 
 
\textit{Lagrangian mechanics.} The CH model sees the cell as a thermodynamic system with two intrinsic degrees of freedom (the two main curvatures defined in terms of their geometrical invariants, $H$ and $K$), and two constraints ($A$ and $V$). A mechanical viewpoint is possible, provided the CH free-energy was viewed as the action that drives the motion of a material particle in the membrane-representing surface $\Omega$ (Fig. 1A) \cite{32}. In the cartoon of Fig. $1A$, such a membrane particle is assumed to travel a trajectory with a kinematics characterized by a set of determined by the local curvature status, which is parameterized in $\Omega$ by the pathway $\boldsymbol{S} (\ell)$ with $ \ell $ being an arbitrary parameter. Thus, the free-energy is interpreted as the action of the material particle along $S(\ell)$, this is \cite{32,55}: 
\begin{equation}
\label{eq. 2}
F_{CH}=2\pi \int d \ell\ \mathfrak{L}_{CH}[\boldsymbol{S} (\ell),\dot{\boldsymbol{S}}(\ell),\ddot{\boldsymbol{S}}(\ell)].
\end{equation}

The global action $F_{CH}$ arises from a Lagrangian density $\mathfrak{L}_{CH}$ with the same units as the free-energy. Because the elastic curvature energy is quadratic on the surface curvatures (Eq. \eqref{eq.1}), the CH-Lagrangian function $\mathfrak{L}_{CH}$ necessarily involves up to second derivatives on the field variables \cite{45,46,56}. The vector tangent to the surface corresponds to the velocity of the particle $\dot{\boldsymbol{S}} = d\boldsymbol{S}/d\ell$, whereas the curvature itself is related to its acceleration, $\ddot{\boldsymbol{S}}=d^2\boldsymbol{S}/d\ell^2$ \cite{56}. The above kinematic interpretation assigns the equilibrium shape to the pathway of least action ($\delta F_{CH}=0$). The formalism has been extensively exploited to generate the equilibrium shapes of membrane vesicles in the axisymmetrical setting \cite{32,55,57,58}.

\textit{Axial symmetry: Cylindrical coordinates.} In the following, we consider rotational symmetry around the $z$-axis, which enables parametrization in cylindrical coordinates. The problem is then reduced to a planar curve in $\mathbb{E}^2$ (Fig. $1B$); this is $ \{ \rho (\ell), z(\ell),\phi=0 \}$ with the radial coordinate running in $\rho \in (0, R]$ ($R$ is the cell radius defined in Fig. $1C$). For the axisymmetric CH-cell, the Lagrangian reads as (see Appendix A): 
\begin{equation}
\label{eq. 3}
\begin{aligned}
&\mathfrak{L}^{(\ell)}_{CH}(\rho ,z,\dot{\rho}, \dot{z},\ddot{\rho}, \ddot{z})= \frac{\kappa}{2}\rho v \left[ \frac{(\dot{\rho}\ddot{z}-\dot{z}\ddot{\rho})}{v^3} + \frac{\dot{z}}{\rho v} - c_0 \right]^2\\
& + \kappa_G\frac{(\dot{\rho}\ddot{z}-\dot{z}\ddot{\rho})\dot{z}}{v^3} + \Sigma \rho v +\frac{P}{2} \rho^2 \dot{z},
\end{aligned}
\end{equation}
which is written as a function of the two spatial coordinates $(\rho, z)$ and their parametric derivatives, $\dot{\rho} = d\rho/d\ell$, $\ddot{\rho} = d^2\rho/d\ell^2$, $\dot{z} = dz/d\ell$ and $\ddot{z} = d^2z/d\ell^2$, with $v(\ell)=\sqrt{\dot{\rho}^2+\dot{z}^2}$ being the modulus of the curve velocity in terms of the {\it geometric time} ($\ell$). Parametric formulas for $A$ and $V$ are deduced in Appendix B in cylindrical coordinates. An equivalent expression to Eq. $(3)$ was obtained previously by Capovilla et al. \cite{56}, for the case $c_0=0$ described under the same parametrization. 

Due to its high dimensionality, the problem here posed is strongly tangled \cite{45,46,56}. However, the determinant of the Hessian matrix is identified null in this case, telling us that the problem is not expressed with independent variables. Indeed, a membrane particle has only one degree of freedom when moving in a linear pathway; consequently, its configuration space should be naturally described by only one generalized coordinate $q$, and its generalized velocity, $\dot{q}$. We tackle thus the question to search for a minimal Lagrangian $L(q, \dot{q}, t)$ able to generate a canonical Hamiltonian $H(q, p, t)$ in generalized coordinates $(q,p)$. 

\section{\label{sec.2} Minimal Canham-Helfrich Lagrangian}

Our first objective consists of searching for a Lagrangian that represents the motion in time of a system-equivalent particle described by a generalized coordinate, {\it i.e.} a functional $L(q,\dot{q},t)$ expressed in terms of the {\it physical time} ($t$), instead of the {\it geometrical time} ($\ell$). Our strategy consists of: $1)$ simplifying by using a reference frame that allows a representation in generalized coordinates; $2)$ reducing possible scale factors, and; $3)$ identifying the minimal Lagrangian with a mechanical analogous. The sequential transformations giving rise to such a Lagrangian reduction are: 

\paragraph{Invariance under time reparametrization.} We will take advantage of the invariance that offers the fluidity property under reparametrizations of the membrane coordinates [49-50]. The functional $\mathfrak{L}_{CH}^{(\ell)}$ is transformed into a Lagrangian that describes dynamics in physical time $(t)$ maintaining the action invariant. Let's $x_i(\ell)=\{\rho(\ell), z(\ell) \}$ the former set of coordinates parametrized by $\ell$, which determines $\mathfrak{L}^{(\ell)}(x_i,\partial_{\ell} x_i, \partial^2_{\ell} x_i)$. Then, we reparametrize $\mathfrak{L}^{(\ell)} \leftrightarrow \mathfrak{L}^{(t)}$ under the transformation $\ell = f(t)$, where $f$ is an arbitrary function of time. Consequently, the surface coordinates can be rewriten as $x_i(\ell) = x_i[f(t)]=y_i(t)$, with partial time derivatives $\partial_{\ell} x_i = (d \ell / d t)^{-1}\partial_t y_i $ and $\partial^2_{\ell} x_i = (d\ell/dt)^{-2}\partial^2_t y_i - (d\ell/ d t)^{-3}(d^2\ell/dt^2) \partial_t y_i $. Without loos of generality, we can ever choose a linear transformation between $\ell$ and $t$ that involves a characteristic rate $\tau^{-1}$. In the simplest case, we take $\ell = t/\tau$, with the rate $d\ell/dt = \tau^{-1}$ defined as a positive constant. After time reparametrization, the Lagrangian $\mathfrak{L}^{(\ell)} $ is invariantly transformed into $\mathfrak{L}^{(t)} (y_i, \partial_t y_i, \partial^2_t y_i)$ through the sequence of transformations:
\begin{widetext}
\begin{equation}
\label{eq. 4}
\begin{aligned}
S \equiv F_{CH}&=\int d\ell \ \mathfrak{L}^{(\ell)} [x_i(\ell), \partial_{\ell}x_i(\ell), \partial^2_{\ell} x_i(\ell) ]  \overset{\ell=f(t)}{=} \int dt\  \mathfrak{L}^{[f(t)]}\left\{ x_i[f(t)], \partial_{\ell} x_i[f(t)], \partial^2_{\ell} x_i[f(t)]\right\} d\ell/ dt \\
&\overset{f(t) \leftrightarrow t/\tau}{=} \tau^{-1} \int dt\ \ \mathfrak{L}^{(t)}[y_i(t), \partial_ty_i(t), \partial^2_t y_i(t)].
\end{aligned}
\end{equation}
\end{widetext}

Since the rate $\tau^{-1}$ relates the geometric parameter $\ell$ with $t$, it can be understood as the scale factor that stablishes the proportionality between them, {\it i.e.} $\ell / \alpha = t / \beta$, with $\alpha$ being the dimensionless period of the path a parameterized in $\ell$, and $\beta$ the period as parameterized in $t$; obviously, $\alpha$ and $\beta$ recapitulate under the rate $\tau^{-1} = \alpha/\beta$. The reparametrization in Eq. \eqref{eq. 4} does not change the form of the Lagrangian, but provides us with a dynamical setting $\mathfrak{L}^{(t)} (y_i, \dot{y}_i, \ddot{y}_i, t)$ given in terms of time-dependent coordinates $y_i(t)$. 

\paragraph{Change of reference system: Galilean invariance.} Now, we adopt a genuine mechanical standpoint, with $\mathfrak{L}_{CH}^{(t)}$ generating the action of a fictitious particle that travels the trajectory represented by the cell profile. As far $\mathfrak{L}_{CH}^{(t)}$ deals with the axisymmetric setting, we can explicit an expression of $\rho(t)$ that allows to track a meridian trajectory $z = z[\rho(t)]$. We take advantage of the Galilean transformation from a static reference frame (laboratory) to an inertial reference system (tracking), which moves along the radial direction (Fig. 1). Specifically, the radial coordinate $\rho(t)$ is transformed in a new coordinate $\rho (t) \rightarrow \rho' (t) = \lambda t$, which varies along the $\rho$-axis at a constant velocity $\lambda$ \cite{23}. From the membrane point-of-view, $\lambda$ is a systemic velocity related to the rate $ \tau^{-1}$; since the radial coordinate covers its full domain $(0, R]$ in a time $\tau$, then $\lambda = R / \tau$ (Fig. 1C). Under the specified Galilean transformation, one gets $z(t) \rightarrow z(t)$ and $\rho(t) \rightarrow \lambda t$, with time-derivatives $\dot{\rho}(t) \equiv \partial_t \rho = \lambda $ and $\ddot{\rho}(t) \equiv \partial^2_t \rho = 0$. Since the Galilean transformation is action-invariant, the CH-Lagrangian is reformulated within the new coordinate system as:
\begin{equation}
\begin{aligned}
\label{eq. 5}
\mathfrak{L}^{(t)}_{CH}(\dot{z}, \ddot{z},t)=&\frac{\kappa}{2} \lambda tv \tau \left[ \frac{\ddot{z}\lambda }{ v^3} + \frac{\dot{z}}{\lambda t v} - c_0 \right]^2 + \kappa_G\frac{ \lambda \ddot{z} \dot{z} \tau}{  v^3}\\
& + \Sigma \lambda t v\tau + \frac{P}{2} \lambda^2 t^2 \dot{z}\tau,
\end{aligned}
\end{equation}
with $v=\sqrt{\lambda^2 + \dot{z}^2}$. This Lagrangian is already minimal (one degree of freedom), as implies only one action coordinate to describe the planar curve in $\mathbb{E}^2$ ($\dot{z}$, and its derivative $\ddot{z}$; see Fig. 1). 

Here, thanks to Galilean transformation, a breakthrough is made in the race towards the canonical Lagrangian; the paradigm has shifted from the static point of view of a CH energy density towards the dynamic description of the moving particle in an energy field. 

\paragraph{Dimensionless variables: Scaling reduction.} We further transform $\mathfrak{L}_{CH}^{(t)}$ to a scale invariant represention using dimensionless variables reduced by systemic parameters: $R$ for the spatial scale, $\tau$ for the time scale and $\lambda =R/\tau$ for the velocities. Specifically, we define $\bar{z} = z/R$, $\bar{c}_0=c_0  R$, $\dot{\bar{z}} = \dot{z}/\lambda$, $\bar{v} = v/\lambda$, $\ddot{\bar{z}} = \ddot{z} \tau/\lambda$ and $\bar{t}=t/\tau$. Because $\kappa$ determines the energy scale, the reduced Lagrangian is defined as $\bar{L}_{CH} = \mathfrak{L}_{CH}/ \kappa$, the dimensionless Guassian susceptibility as $\chi_G = \kappa_G/\kappa$, and the reduced membrane tension and external pressure as $\bar{\Sigma} = \Sigma R^2/\kappa$ and $\bar{P} = P R^3/\kappa$, respectively. Finally, to get the Lagrangian with the canonical form $\bar{L} (q,\dot{q},\bar{t})$, we make the change of variable $q: = \dot{\bar{z}}$, then $\dot{q}:=dq/d \bar{t}=\ddot{\bar{z}}$, with the time derivative defined in terms of the reduced time $\bar{t}$. Therefore, under the definition $\bar{L}_{CH} (q,\dot{q},\bar{t}) := \mathfrak{L}^{(t)}_{CH} (\dot{z}/\lambda, \ddot{z} \tau/\lambda, t/\tau)/ \kappa$, one obtains:
\begin{equation}
\label{eq. 6}
\begin{aligned}
\bar{L}_{CH} (q, \dot{q},\bar{t}) = \frac{\bar{t} \bar{v}}{2} \left[ \frac{ \dot{q} }{\bar{v}^3} + \frac{ q }{\bar{t} \bar{v}} - \bar{c}_0 \right]^2 + \chi_G \frac{q \dot{q} }{\bar{v}^3} + \bar{L}_0,
\end{aligned}
\end{equation}
with $\bar{L}_0 = \bar{\Sigma} \bar{t} \bar{v} + \bar{P} \bar{t}^2 q/2$. This Lagrangian already represents the motion (in one-dimension) of a fictitious particle, characterized by the generalized coordinate $q$, and the generalized velocity $\dot{q}$.

\paragraph{Spring-mass equivalent.} The mechanical system represented by Eq. (6) actually corresponds to the curvature motions of the meridian profile (Fig. 1). However, the material content of such cell is not considered by $\bar{L}_{CH}$, which actually corresponds to a massless particle. To take adequate account of translational motions, we assume the elastic entity described by Eq. (6) to be coupled to a membrane-equivalent particle with intrinsic mass $M$. Figure $2A$ depicts the mechanical equivalent as a spring-mass system described in terms of the generalized coordinate $q$ (representing the inclination of the curve relative to $\lambda$, {\it i.e.} $q = \dot{z}/\lambda$) and of the generalized velocity $\dot{q} = dq/d\bar{t}$ (representing its reduced curvature in the vertical direction, {\it i.e.} $\dot{q} = \ddot{z} \tau/\lambda$). As regards the generalized moment $p = \partial \bar{L}_{CH}/\partial \dot{q}$, it represents the curvature stress $\sigma_{\ddot{z}} = \ \partial \mathfrak{L}_{CH}/\partial \ddot{z} = (\tau/\lambda) \ \partial L_{CH}/\partial \dot{q} = R \mu_0 p$, which we will refer to as $S \equiv \sigma_{\ddot{z}}$, as corresponds to changes in curvature. Here, the mass $\mu_0 = \kappa (\tau / R)^2$ determines the scaling factor between the dimensionless moment $p$ and the moment of curvature $S = I_1 p$ with the units of the first moment of mass, this is $I_1 = \mu_0 R$. The elemental mass $\mu_0$ represents an effective mass that depends on the local rigidity of the membrane $\kappa$. Specifically, $\mu_0$ is the inertial mass that interacts with the curvature-elasticity field, which causes an elastic response in the membrane-equivalent entity as internal ``vibrational modes" (Fig. $2B$). In principle, such a spring-equivalent inertial mass $(\mu_0)$ is different to the material mass of the membrane $(M)$; in terms of inertia, $M$ represents how ``heavy" the membrane behaves against motion, and $\mu_0$ how ``rigid" is the equivalent particle-spring interaction.

 \begin{figure}
 \centering
 \includegraphics[width=8.5cm,height=6.5cm]{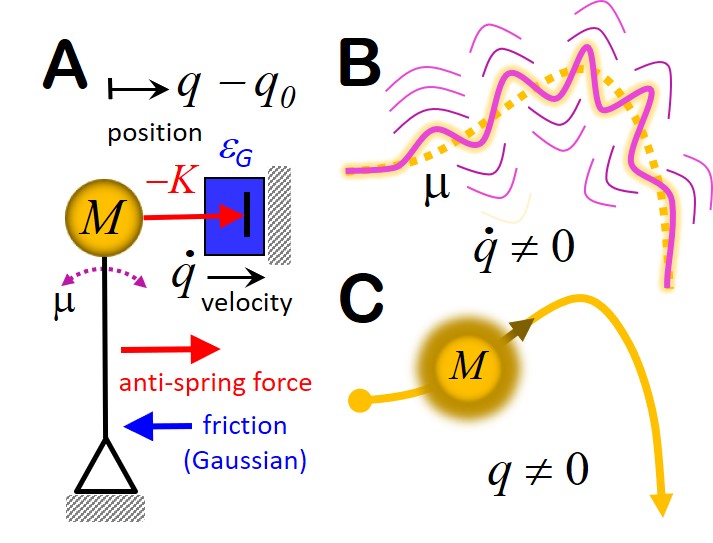}
 \caption[Texto para la lista]{ $A)$ Spring-mass equivalent of a cellular profile described as a membrane-equivalent material particle interacting with the Canham-Helfrich curvature-elasticity field. This one-dimensional mechanical system is kinematically described by the normal coordinate $q$ (representing the inclination of the cellular profile), its relative displacement with respect to the equilibrium configuration $q_0$ (representing the natural inclination of the curve due to spontaneous curvature), and the velocity $\dot{q}$ (representing the local curvature as an instantaneous change in inclination). The elastic element is characterized by: $i)$ effective anti-spring constant $K (= \kappa/\bar{v} \bar{t})$, accounting for the traction force of the inverted oscillator; $ii)$ time-dependent inertial mass $\mu (= \mu_0 \bar{t}/\bar{v}^5)$, accounting for the effective mass sensed by the CH-field (with rest inertial mass $\mu_0 = \kappa/\lambda^2)$; and $iii)$ Gaussian permitivity $\varepsilon_G (= 1+\kappa_G/\kappa)$, accounting for effective dissipation as viscous drag against motion. The material element is explicitely described by the invariant intrinisic mass $M$. Two dynamical modes are present within this dissipative spring-mass system: $B)$ Vibrational mode corresponding to the oscillation of the curve that describes the cell profile; this mode is due to elasticity thus implies a curvature ($\dot{q} \neq 0$), which is equivalent to the oscillation of the inverted oscillator of apparent mass $\mu$. $C)$ Traslational mode of the membrane equivalent particle of intrinisic mass $M$; this kinetical mode describes the translational path necessary to track the material profile of the cell.}
 \label{fig.2}
 \end{figure}

\section{\label{sec.3} Total Lagrangian}

With these considerations in mind, the CH-Lagrangian is reorganized with the form $L = T - U$, where $T$ is the kinetic energy and $U = V + Q + C$, a generalized potential, with $V$ being the potential energy, $Q$ a dissipative function and $C$ an independet term due to constraints. 

\textit{Inertial mass: Vibrational kinetic energy.} Becase Eq. \eqref{eq. 6} represents a spring-equivalent particle of rest mass $\mu_0$ with velocity $\dot{q}$, its kinetic energy would be proportional to $\mu_0$ and quadratic on $\dot{q}$. After expanding, the kinetic term is identified  as:
\begin{equation}
\begin{aligned}
\label{eq. 7}
T_{CH} & = \frac{1}{2} \mu \lambda^2 (\dot{q} - \bar{c}_0 \bar{v}^3)^2,
\end{aligned}
\end{equation}
which has units of energy, with a time-dependent effective mass defined as $\mu = \bar{\gamma} \mu_0 $.  Here, the dynamic factor $\bar{\gamma} = \bar{t}/\bar{v}^5$ takes a similar role as the Lorentz-factor in special relativity, renormalizing the mass at rest $\mu_0$ by a factor that depends on the velocity. In this case, the faster the motion the lighter the particle; indeed, an effectively massless regime is attained at high velocity ($\bar{\gamma} \rightarrow 0$ if $\bar{v} \rightarrow \infty$). Because the kinetic energy is essentially field-dependent (exclusively arising from changes in curvature, {\it i.e.} vibrational modes), an additional kinetic component representing the translational energy of the material constituents might be consistently considered. 

\textit{Material mass: Translational motion and total kinetic energy.} Let's consider the membrane-equivalent particle to bear an intrinsic mass $M$ moving along the membrane-representing curve (Fig. $2C$). The translational kinetic energy is $T_{trans} = M (\dot{x}^2 + \dot{y}^2 + \dot{z}^2) / 2$, or in terms of the generalized coordinate $T_{trans} (M, q) = M \lambda^2 (1 + q^2) / 2 $. Consequently, the total kinetic energy stands on $T = T_{CH} (\mu, \dot{q}) + T_{trans} (M, q)$, with the curvature-dependent component $T_{CH}$ proportional to the effective inertial mass of the equivalent spring $\mu$ (vibrational mode; Fig. $2B$), and the inclination-dependent $T_{trans}$ proportional to the material mass $M$ (translational mode; Fig. $2C$). The relationship between the two masses is given by the dimensionless number $\eta = \mu_0 / M$, which demarcates three different inertial regimes: {\it i)} Rigid membrane with effective mass governed by bending stiffness $(\eta >> 1)$, for which the kinetic energy is dominated by elastic vibrational modes, {\it i.e.} $T \approx T_{vib} \sim \mu \dot{q}^2 $. {\it ii)} Flexible membrane with effective mass governed by its intrinsic mass ($ \eta << 1 $),  with a kinetics dominated by inertia, {\it i.e.} $T \approx T_{trans} \sim M q^2$. {\it iii)} Equivalence between stiffness and inertia ($\eta = 1 $), which implies $\mu_0 = M $. 

\textit{Potential energy: Mean curvature.} The $q$-dependent terms in the Lagrangian can be considered as a potential energy describing the mass-spring elastic interaction. The potential energy $V$ is a repulsive parabolic barrier:
\begin{equation}
\label{eq. 8}
V(q, \bar{t}) = -\frac{1}{2} K (q - q_0)^2,
\end{equation}
which is centered at $q_0 = \bar{c}_0 \bar{v} \bar{t}$, with stiffness $K = \kappa / \bar{v} \bar{t}$. This potential represents the force field sensed by the membrane-equivalent particle to create curvature in configuration-space. The anti-spring driving force is given by the effective elastic stress caused by the $CH$-elasticity field, this is $f_{elas} = - \partial V/\partial q = K (q - q_0) $. 

The time-dependent stiffness $K = \kappa/ \bar{\ell}_{\lambda}$ is determined by the reduced distance $\bar{\ell}_{\lambda} = \bar{v} \bar{t}$, which represents the configurational pathlength traveled by the particle after a time $\bar{t}$; it modulates both the equilibrium coordinate ($q_0 = \bar{c}_0 \bar{\ell}_{\lambda}$) and the effective stiffness of the repulsive barrier ($K/\kappa = \bar{\ell}_{\lambda}^{-1}$). Note that the effective stiffness diverges at short times ($K \rightarrow \infty$ at $\bar{t} \rightarrow 0$), which makes the particle to be repeled out from the metastable configuration at $q_0$. However, as time proceeds the barrier softens ($ K \rightarrow 0$ at $\bar{t} \rightarrow \infty$), which allows the system to reach steady-state configurations far away from $q_0$. 

\textit{Dissipation: Gaussian friction}. Drag forces arise from dissipative potentials that depend on the velocity. All terms on the crossed product $q\dot{q}$ in Eq. (6) are regrouped as a dissipative function due to Gaussian elasticity:  
\begin{equation}
\label{eq. 9}
Q (q, \dot{q})/\kappa = -  \varepsilon_G  \frac{q \dot{q}}{\bar{v}^3},
\end{equation}
which accounts for a shearing force proportional to curve inclination with a strength given by the permitivity $\varepsilon_G = 1 + \chi_G = 1 + \kappa_G/\kappa$ (Gaussian shear involves a reactive coupling between curvature and inclination; see Eq. $A8$). 

This dissipative function actually describes a drag force $f_{frict} = d / dt (\partial Q / \partial \dot{q}) = - \xi_G \dot{q}$, which opposes to movement. The effective friction coefficient is defined as $\xi_G = \kappa \varepsilon_G/\bar{v}^3$, being constitutively determined by the Gaussian permitivity $\varepsilon_G$. Whereas $\varepsilon_G$ measures how much energy must be spent upon creating Gaussian curvature, its friction-equivalent $\bar{\xi}_G (=\xi_G/\kappa)$ determines the resistance of the spring-mass system against creating curvature at a velocity $\dot{q}$ (Fig. $2A$). Upon this definition, different frictional regimes can be distinguished: 

{\it i)} Viscous drag ($\bar{\xi}_G > 0$) at positive Gaussian permitivity ($\varepsilon_G > 0$), representing a dissipative dynamics characterized by a relatively low saddle-splay rigidity ($\chi_G > -1$; $\varepsilon_G > 0$). Such a status of Gaussian softness results in a significant viscous drag in the equivalent mechanical system, which implies a frictional resistance to create mean curvature at the expense of a facilitated tendency to create Gaussian curvature. 

{\it ii)} Frictionless ($\bar{\xi}_G = 0$), which corresponds to typical lipid bilayers ($\chi_G \approx - 1$; $\varepsilon_G \approx 0$). In this case, the shear strain field is cancelled out resulting in effective null friction. In practice, no strong influence of the dissipative function is expected thus in real biomembranes. 

{\it iii)} Propulsion ($\bar{\xi}_G < 0$), which represents hypothetical situations with a negative Gaussian permitivity ($\varepsilon_G < 0$), corresponding to very high values of the Gaussian stiffness ($\chi_G < -1 $). In these cases, creation of Gaussian curvature is highly impeded, resulting in a coupling that favours curvature creation at the expense of the bending component of the curvature-elasticity field. 

\textit{Constraints.} Finally, the independent term $C$ due to constraints is identified as:
\begin{equation}
\label{eq. 10}
C(q, \bar{t})/\kappa = \frac{\bar{\ell}_{\lambda}}{2} \bar{c}_0^2 - \bar{L}_0,
\end{equation}
where the configurational length $\bar{\ell}_{\lambda}$, enforced by the quadratic spontaneous curvature $(\bar{c}_0^2)$, opposses to the systemic constraints defined by $\bar{L}_0$. The constraint function $C (q, \bar{t})$ is holonomic (it does not depend on $\dot{q}$, or any higher order time derivatives), thus, the final steady-state of the constrained status will not depend on the intermediate values of the trajectory. The holonomic constraint $C( q, \bar{t}) = 0$, which represents equilibrium betweeen constraining forces stablishes a strong condition for the configurational coordinate; this is the mechanical statement of the Young-Laplace equation.

\textit{Total Lagrangian: Reduced form.} The mechanical equivalent of the cell membrane has been identified as an inverted mass-spring oscillator of constant $- K$ and friction $\xi_G$, which produces damped motion in a membrane-equivalent particle of material mass $M$ and inertial mass $\mu$ (Fig. $2A$). Once added the translational term to the kinetic energy $T = T_{CH} + M \lambda^2 (1+q^2)/2$, and considered the components of the generalized potential $U (= V + Q + C)$, two elemental energies can be identified, respectively, as the amplitudes of the kinetic ($T_0 = M \lambda^2 / 2$) and potential ($U_0 = \kappa / 2$) terms. Both are mutually related as $T_0 \eta = U_0$ (at equivalence $\eta = 1$, thus $T_0 = U_0$, describing the particular case when the elastic energy of the membrane equals the kinetic energy of its material content). Therefore, a general reduced Lagrangian form is possible in terms of the elemental energy $\kappa$; under the definition $\bar{L} := L/\kappa = (T - U)/\kappa$, one gets: 
\begin{widetext}
\begin{equation}
\label{eq. 11}
\bar{L}(q,\dot{q},\bar{t}) = \bar{T} - \bar{U} =  \left[ \frac{\bar{\gamma}}{2} \left( \dot{q} - \dot{q}_0 \right)^2 + \frac{1}{2 \eta} \left( 1 + q^2 \right) \right]  -  \left[ - \frac{\bar{K}}{2} (q - q_0)^2 - \varepsilon_G \frac{q \dot{q}}{\bar{v}^3} + \frac{1}{2}\bar{\ell}_{\lambda}\bar{c}_0^2 - \bar{L}_0 \right],
\end{equation}
\end{widetext}
where the two separated components ($\bar{T}$ and $\bar{U}$) are written in terms of dimensionless dynamical parameters $\bar{\gamma} = \bar{t}/\bar{v}^5$, $\bar{\ell}_{\lambda} = \bar{v} \bar{t}$, viscoelastic parameters $\bar{K} = K/\kappa = \bar{\ell}_{\lambda}^{-1}$ and $\bar{\xi}_G = \varepsilon_G / \bar{v}^3$, mass parameters $\eta = \mu_0/M$, and configurational coordinates $q_0 = \bar{c}_0 \bar{\ell}_{\lambda}$ and $\dot{q}_0 = \bar{c}_0 \bar{v}^3 = q_0^2 (\bar{v}^2/\bar{t})$. The independent term $\bar{L}_0$ due to area/volume constraints was previously defined in Eq. (6).

\textit{Canonical moment.} In the way to the canonical Hamiltonian, we recall again on the generalized moment $(p)$, which is conjugated to the generalized coordinate $(q)$. In terms of the reduced Lagrangian $\bar{L}$, one gets: 
\begin{equation}
\label{eq. 12}
\begin{aligned}
p (q, \dot{q}, \bar{t}) = \frac{\partial \bar{L}}{\partial \dot{q}} =  p_0 +\bar{\gamma} \dot{q} + \varepsilon_G \frac{q}{\bar{v}^3},
\end{aligned}
\end{equation}
with equilibrium value $p_0 = - \bar{c}_0 (\bar{t}/\bar{v}^2) = - \bar{\gamma} \dot{q}_0$. 

As stated above, the generalized moment $p$ represents how the curvature configurates to define the field of curvature stress with respect to the first moment of inertia $I_1 = \mu_0 R$; this is $ S = \partial \mathfrak{L}_{CH}/\partial \ddot{z} =  I_1 p$, thus, $p := S /I_1$. High curvatures (with respect to $I_1$) are represented by $p > 1$, which correspond in general to high configurational velocities $\dot{q}$. Conversely, $p < 1$ corresponds to low curvatures. For very flexible membranes ($\kappa \rightarrow 0$), both $\mu_0 \rightarrow 0$ and $I_1 \rightarrow 0$; thus, the generalized moment is defined to be zero in this limit ($p \rightarrow 0$). The generalized moment defined by Eq. \eqref{eq. 12} can be decomposed in three terms as $p = p_0 + p_{mean} + p_{Gauss}$. The first term $p_0$ imposes the equilibrium status in phase space, with coordinates completely determined by the spontaneous curvature $q_0 = \bar{c}_0 \bar{\ell}_{\lambda}, \dot{q}_0 = \bar{c}_0 \bar{v}^3, p_0 = -\bar{\gamma} \dot{q}_0$ and $p_0 \dot{q}_0 = - \bar{c}_0^2 \bar{\ell}_{\lambda}$; if $\bar{c}_0 = 0$, then $q_0 = 0$ and $\dot{q}_0 = 0, p_0 = 0$. The configurational term due to changes in mean curvature $p_{mean} = \bar{\gamma} \dot{q}$, actually corresponds to the linear momentum of the membrane-equivalent particle, {\it i.e.} $p_q = \mu_0 p_{mean} = \mu \dot{q}$. Finally, the term $p_{Gauss} = (\varepsilon_G/\bar{v}^3) q $ represents the configurational contribution from Gaussian curvature to the curvature momentum.

\section{\label{sec.3} Canonical Hamiltonian}

Once a dimensionally minimal Lagrangian has been obtained in terms of generalized variables $ L = \kappa \bar{L} (q, \dot{q}, \bar{t})$, the Legendre transformation enables getting the equivalent Hamiltonian $H(q, p, t)$ \cite{21}. Specifically, we recall on the reduced form $\bar{H}(q, p,\bar{t})=\dot{q}p - \bar{L}(q,\dot{q},\bar{t})$, with $\bar{H} = H/\kappa$. After analytics (see Appendix D), the canonical Hamiltonian is obtained as the sum of four additive terms as:
\begin{equation}
\label{eq. 13}
\bar{H}(q, p, \bar{t}) = \bar{H}_{kin} + \bar{H}_{pot} + \bar{H}_{frict} + \bar{H}_C,
\end{equation}
which are identified as a kinetic function $\bar{H}_{kin}$, a pure-potential term $\bar{H}_{pot}$, a friction component $\bar{H}_{frict}$ and  the independent term $\bar{H}_C$ due to constraints.

\textit{Kinetic function.} The kinetic component $\bar{H}_{kin}$ is obtained as the difference between two quadratic terms:
\begin{equation}
\label{eq. 14}
\bar{H}_{kin} (p, q, \bar{t}) =  \bar{H}_{kin}^{(vib)} - \bar{H}_{kin}^{(trans)} = \frac{(p-p_0)^2}{2 \bar{\gamma}} - \frac{1 + q^2}{2 \eta}.
\end{equation}

The kinetic energy is not only a function of the generalized moment, but also of the coordinate and time (notice that $\bar{\gamma} = \bar{t}/(1+q^2)^{5/2}$). This makes $\bar{H}_{kin} (p, q, \bar{t})$ a configurational property giving rise to non-conservative forces ($\partial \bar{H}_{kin}/\partial q \neq 0$). Whereas the vibrational component $\bar{H}_{kin}^{(vib)}$ is a genuine elasticity-field contribution, as far it defines the kinetic energy of the curvature motions, the translational term $\bar{H}_{kin}^{(trans)} ( = \bar{T}_{trans})$ exclusively corresponds to the displacement of the membrane as a whole. Because the kinetic status is described by $p$ but not $q$, thus, the difference $\bar{H}_{vib} - \bar{H}_{trans}$ represents the kinetic energy available to change the curvature configuration after the translational energy is substracted thereof \cite{62}. 

\textit{Potential function.}  Under the definition of the configurational energy possessed by the particle due to its generalized position in the curvature-elasticity field $(q)$, two additive potential contributions are identified as:
\begin{equation}
\begin{aligned}
\label{eq. 15}
\bar{H}_{pot} (q, \bar{t}) =& \bar{H}_{pot}^{(mean)} + \bar{H}_{pot}^{(Gauss)}=\\
 & - \frac{1}{2} \bar{K} (q-q_0)^2 + \frac{1}{2} \bar{K} \varepsilon_G^2 q^2,
\end{aligned}
\end{equation}
which correspond, respectively, to: {\it i)} $\bar{H}_{pot}^{(mean)}$ a time-dependent repulsive barrier of strength $\bar{K}$ due to the propensity of the field to create mean curvature (see Eq. (8)); {\it ii)} $\bar{H}_{pot}^{(Gauss)}$ an attractive potential arising from Gaussian elasticity. Both fields are governed by the effective strength $\bar{K} \equiv \bar{\ell}_{\lambda}^{-1}$ and extend out the configurational space a length $\bar{\ell}_{\lambda} (= \bar{v} \bar{t})$, which is completely determined by the kinematical status. 

 \begin{figure*}
 \centering
 \includegraphics[width=17cm,height=10.5cm]{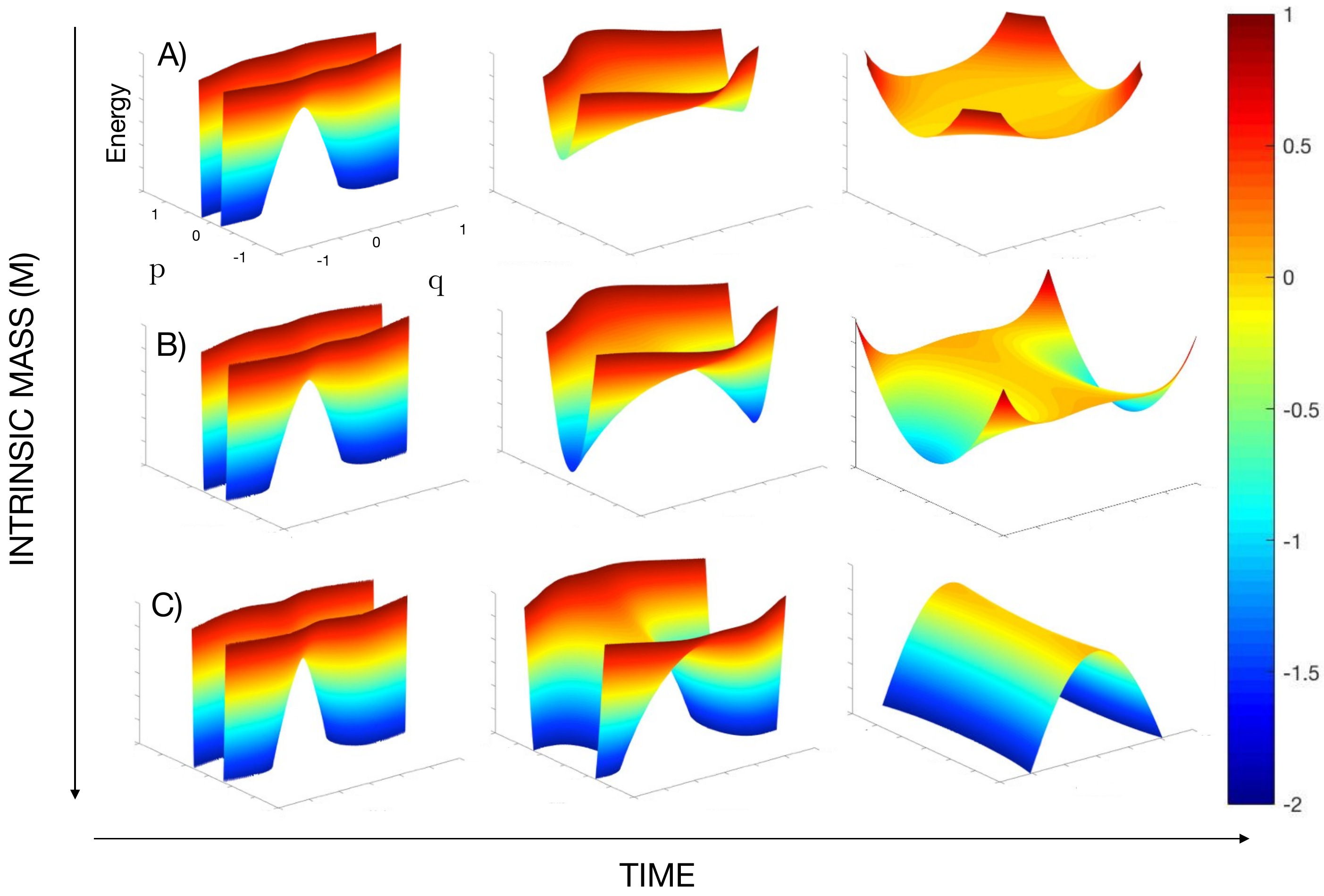}
 \caption[Texto para la lista]{Time evolution of the energy landscape in the frictionless case of a lipid bilayer membrane ($\chi_G = -1; \varepsilon_G = 0$; $\xi_G = 0$) (from left to right) for inertial status: $A)$ Massless, predominantly rigid, membrane ($M = 0; \eta \rightarrow \infty$). $B)$ Mass equivalence ($M = \mu_0 ; \eta = 1$). $C)$ Flexible membrane with a comparatively high intrinsic mass ($M >> \mu_0; \eta << 1$).}
 \label{fig.3}
 \end{figure*}

\textit{Friction: Power function.} Additionally, the Gaussian rigidity contributes the generalized potential with a $p$-dependent frictional term:
\begin{equation}
\label{eq. 16}
\bar{H}_{frict} (p, q, \bar{t}) = - \bar{\xi}_G \frac{q}{\bar{\gamma}} (p - p_0) ,
\end{equation}
which arises from the dissipative function $(Q)$ defined in Eq. (9). This dissipative potential is proportional to the effective friction coefficient due to Gaussian stiffness $\bar{\xi}_G (= \varepsilon_G/\bar{v}^3)$. Because frictional drag decreases the energy available to the membrane-equivalent particle, the negative contribution $\bar{H}_{frict}$ is interpreted as a fricitional energy dissipated upon creating curvature. Indeed, it opposses to motion varying as the first power of the generalized moment, two conditions necessary to be identified as a frictional drag. A generalized power function includes all the potential terms with a dissipative nature; in reduced form, this is $\bar{\Pi} (p, q, \bar{t}) = \bar{H}_{pot} (q, \bar{t}) + \bar{H}_{frict} (p, q, \bar{t})$. The power function $\bar{\Pi}$ is analogous to a generalized potential function, but broader in scope as frictional forces are made explicit.

\textit{Constraints.} The independent term $\bar{H}_C$ represents the virtual work due to the preservation of the systemic constraints. This term is specified by the constraint function $\bar{C} (q, \bar{t}) = C (q, \bar{t})/ \kappa$ (see Eq. \eqref{eq. 10}); in reduced form: 
\begin{equation}
\label{eq. 17}
\bar{H}_{C} (q, \bar{t}) = -  \frac{1}{2} \bar{\ell}_{\lambda} \left( 2 \bar{\Sigma} + \bar{P} \frac{\bar{t}}{\bar{v}} q \right),
\end{equation}
which determines the time-dependent energy of the vacuum state (undeformed), this is $E_0 = \kappa \bar{H}_C (q, \bar{t})$. 

\textit{Total energy: Non-conservative dynamics.} The complete Hamiltonian $\bar{H} = \bar{H}_{kin} + \bar{\Pi} + \bar{H}_C$ is rehonomic as contains the time as an explicit variable, {\it i.e.} $\partial \bar{H}/\partial \bar{t} \neq 0$. The function $H = \kappa \bar{H}$ recapitulates the total energy of the minimal cell under the form of an equivalent inverted oscillator with a highly non-conservative character. The so-defined Hamiltonian origins in a particle-field interaction characterised by time-dependent effective inertial mass $\mu$ (giving rise to kinetically non-conserved curvature momentum), time-dependent effective rigidity $K$ (producing elastic forces decreasing with time) and material mass $M$ (describing translational inertia). At short times, the equivalent mass-spring system behaves massless with a dynamics dominated by the curvature-elasticity field ($\mu \rightarrow 0$ and $K \rightarrow \infty$ at $\bar{t} \rightarrow 0$). Conversely, rigidity effects become weaker at longer times, with inertial effects dominating at the end of the trajectory ($\mu \rightarrow \infty$ and $K \rightarrow 0$ at $\bar{t} \rightarrow \infty$). Consequently, the above Hamiltonian is highly non-conservative; $ d\bar{H}/d\bar{t} =  \partial \bar{H}/ \partial \bar{t} + (\partial \bar{H}/ \partial q) \dot{q} + (\partial \bar{H}/\partial p) \dot{p} \neq 0$, specially at short times. At long times, however, it becomes stationary, {\it i.e.} $d \bar{H}/ d \bar{t} = 0$, as expected for steady-state equilibrium conditions. Figure 3 shows the singular time-dependence of the Hamitonian surface for the ideal (frictionless) case. Although near-universal inflactionary behavior is observed at short times, a rich evolution dynamics is expected depending on the dynamic balance between curvature (vibration) and inertia (translation). The history of the energy landscape is narrated as follows:

{\it 1.} {\it Initial inflation.} In the very begining $(\bar{t} \rightarrow 0)$, the energy surface forms around a highly-curved saddle-point centred at the origin of the phase-space $(q_0, p_0)$. As time proceeds, the nascent surface develops two deep-wells along the $q$-axis, which favor quick repulsion from the configurational center $(q_0, p_0)$ towards the outer rim of the repulsive barrier (at high generalized position $|q| >> q_0$, but still low moment $p \approx p_0$; see Fig. 3; left panels). Such a primigenial evolution can be said, in the language of local curvatures, as the result to minimize the curvature ($\ddot{z} \sim \dot{q} \sim p$) at the expense of maximizing the curve inclination $(\dot{z} \sim q)$. Such a universality class of dynamic behavior stems on the repulsive nature of the inverted potential  function $\bar{V} = - \bar{K} (q - q_0)^2 / 2$.
 
{\it 2.} \textit{Relaxation.} This transient stage is characterized by the progressive relaxation of the repulsive barrier ($\bar{K} \sim 1/\bar{t}$). During this stage, the energy landscape develops a metastable plateau around the configurational center at $(p_0, q_0)$ (Fig. 3; central panels). The nascent stability region extends out at larger $p$'s than in the very begining (higher curvatures). The size of this region depends on the configurational kinetic energy available to the system as determined by $M$; the lighter the membrane, the higher the kinetic energy available to develop curvature. 

{\it 3.} \textit{Steady state.} The relaxing energy surface is observed to later evolve towards steady-state state  (Fig. 3; right panels). The final stationary landscape depends strongly on the systemic properties, particularly on $\eta (= \mu_0/M$), defining three material regimes: {\it i)} Rigid ($M << \mu_0$, $\eta >> 1$), for which the transitory kinetic barrier is observed to rapidly vanish under the dominance of membrane stiffness. In this massless regime, $H$ describes an initially repulsive system that evolves towards bounded states (vibrational modes) governed by an attractive effective potential (Fig. $3A$). {\it ii)} $M \approx \mu_0$, which define intermediate cases of material membranes with an intrinsic mass comparable to the inertial mass of the equivalent spring ($\eta \approx 1$); here, both features appear combined together (repulsive/attractive). Particularly, at mass-equivalence ($M = \mu_0$, $\eta = 1$; Fig. 3B), the energy surface is hybrid behaving essentially repulsive along the $q$-axis (exciting translational modes, thus creating inclination) and attractive along the $p$-axis (inhibiting high-curvature modes, thus reducing curvature). Next, we will only elaborate on this particular hybrid case ($\eta = 1$), which represents constitutive conditions of material equivalence between intrinsic mass and rigidity. {\it iii)} Massive membranes ($M >> \mu_0$, $\eta << 1$), where the translational energy dominates, thus making the energy landscape to develop a steady-state repulsive barrier (Fig. $3C$). Such a status gives rise to unbonded states (translational modes) that effectively favour a high inclination at practically any curvature. 

\textit{Symmetries.} Prior to examine the phase-space, we discuss its underlying symmetries through the generating Hamiltonian function. Figure 4 shows the planar countor plots of the Hamiltonian surface for different parametric combinations, with the isoenergetic lines revealing its symmetries. The reference case ($\varepsilon_G = c_0 = \Sigma = P = 0)$ is plotted in the central panel (Fig. 4E). As expected, the configurational energy is quadratic on $q$ and $p$ with its minimum centered at $(q_0, p_0)$. The $p2q2$-symmetry is elongated by the translational component to the kinetic energy, which introduces progressive $q^2$-character; the higher $M$ the dominant the $q2$-symmetry. In general, inversion symmetries ($q \rightarrow - q$ and $p \rightarrow - p$) are imposed by the configurational axes, which splits the configurational space in four sectors with a centrosymmetric structure (I, II, III and IV, counterclockwise; see Fig. 4E). The Gaussian friction is the control factor for breaking biaxial symmetry through the power function; at $c_0 = 0$, one founds $- 2 \bar{\Pi}  \bar{\ell}_{\lambda} \approx (1 -\varepsilon_G^2)q^2 + 2 \bar{\xi}_G \bar{v}^6 qp$. Although the symmetry of the power field is parabolic along the $q$-axis, its inclination is determined by Gaussian friction in a strongly $\bar{v}^6$-dependent fashion.

 \begin{figure}
 \includegraphics[width=8.5cm,height=7cm]{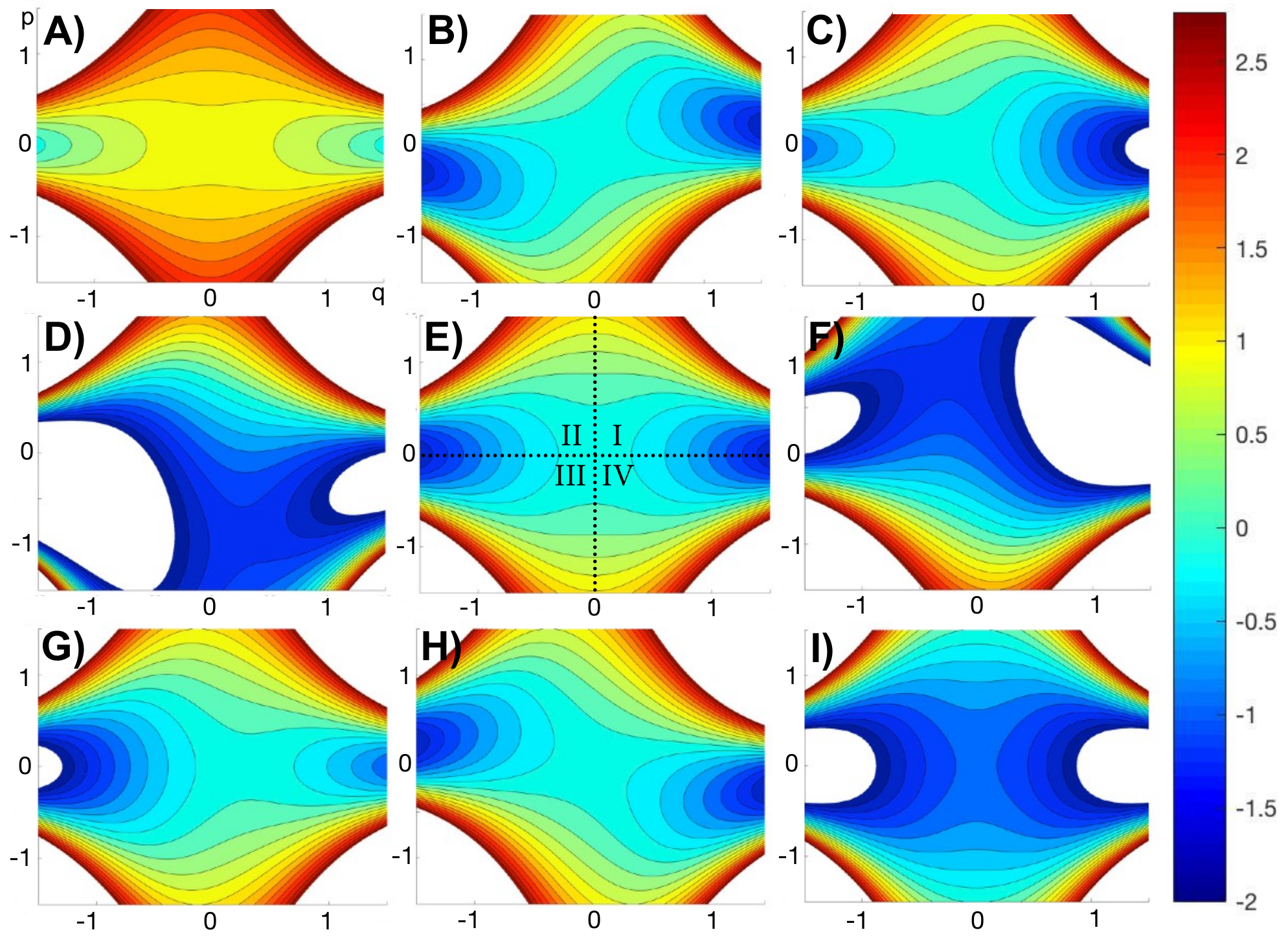}
 \caption[Texto para la lista]{Contour plots of the Hamiltonian surface described through Eqs. (13)-(17) (considered at mass equivalence $\eta = 1$, and at time $\bar{t} = 1$). The reference case of an unconstrained membrane ($\bar{\Sigma} = \bar{P} = \bar{c}_0 = 0$) in the absence of Gaussian curvature effects ($\varepsilon_G =0$) is represented in the central panel (E). This corresponds to a saddle surface built around the configurational center ($H = 0$ at $q_0 = p_0 = 0$), with a $q^2$-barrier (potential) and $p^2$-confinement (kinetic). Here ($p2q2$ symmetry), energy is simetrically distributed between the four equivalent sectors: I $(q > 0, p > 0)$; II $(q < 0, p > 0)$; III $(q < 0, p < 0)$; $IV (q >0, p < 0)$. The presence of finite Gaussian curvature introduces symmetry breaking (vertical central panels); depending on the Gaussian permitivity: B) $\varepsilon_G > 0$, $(+)pq$-symmetry favouring even sectors with negative energies ($H < 0$ at I and III); E) $\varepsilon_G = 0$, symmetric case ($p2q2$); H) $\varepsilon_ G < 0$, $(-)pq$-symmetry favouring odd sectors ($H < 0$ at II and IV). Spontaneous curvature also introduces symmetry breaking (horizontal central panels): D) $\bar{c}_0 > 0$ leading to $p_0 (= - \bar{c}_0 \bar{v}^2 / \bar{t}) < 0$ favours regions with negative value of the curvature moment ($H < 0$ at III and IV); E) $\bar{c}_0 = 0$ (symmetric case); F) $\bar{c}_0 < 0 (p_0 > 0)$ favours positive moments ($H < 0$ at I and II). The cell pressure causes asymmetric $q$-imbalance of the $q2$-symmetry through the volume constraint ($\bar{H}_C^{(V)} = - (\bar{P} \bar{t}^2/ 2) q$; panels in the main diagonal): C) $\bar{P} > 0$ (inflated cell; hipotonic conditions) favours negative energy at $q > 0$ ($H < 0$ at I and IV); E) $\bar{P} = 0$; symmetric (isotonic); G) $\bar{P} < 0$ (deflated cell; hypertonic) favours negative energy at $q < 0$ ($H < 0$ at II and III). The area constraint ($\bar{H}_C^{(A)} = - \bar{\ell}_{\lambda} \bar{\Sigma}$), does not break the $p2q2$-symmetry of the final steady state (negative diagonal): A) $\bar{\Sigma} < 0$ (floppy membrane); depending on the membrane tension: E) $\bar{\Sigma} = 0$ (tensionless); I) $\bar{\Sigma} > 0$ (tensioned membrane).}
 \label{fig.4}
 \end{figure}

Broken symmetries correspond to different material classes characterized by finite Gaussian permitivity, associated to different frictional regimes: {\it i)} Membranes with a spontaneous tendency to create Gaussian curvature ($\chi_G > 0$, $\varepsilon_G > 1$), which represent the typical frictional status $(\bar{\xi}_G > 1)$. In this regime, the power function works as a repulsive driver with an essential frictional character; here, the $q2$-symmetry is broken by $qp$-coupling, which imposes an elliptically deformed symmetry of positive incline determined by the strenght of the frictional force (between sectors I-III; see Fig. 4B). {\it ii)} Ideal membranes (frictionless) with a finite Gaussian stiffness, ($\chi_G \approx - 1$, thus $\varepsilon_G \approx 0$), for which creating Gaussian curvature requires energy input but not sufficient to produce effective friction. In the frictionless regime $(\bar{\xi}_G \approx 0)$, we essentially found a $q2$-symmetry as far the power function is dominated by the pure-repulsive barrier; this is, $\bar{\Pi} \approx - \bar{K} q^2/2$ (between sectors I-II and III-IV; see Fig. 4E). {\it iii)} Membranes with a high Gaussian stiffness ($\chi_G < - 1$, $\varepsilon_G < 0$). This represents the propulsion status able to create additional mean curvature at the expense of a reactive resistence to create Gaussian curvature ($\bar{\xi}_G < 0$). In this case, the structure of the asymmetric break-up adopts a negative incline, which represents the strenght of the Gaussian-mediated propelling force (between sectors II-IV; see Fig. 4H). 

With respect to spontaneous curvature $\bar{c}_0$, a very significant symmetry breaking is induced thereby (Fig. 4D-F). In the cases with $c_0 \neq 0$, the two attractive wells become strongly deformed, sifthing the configurational center at the position $q_0 = \bar{c}_0 \bar{\ell}_{\lambda}$ and $p_0 = - \bar{c}_0 \bar{\ell}_{\lambda} /\bar{v}^3$. The cell pressure induces a similar break-up, building the two attractive wells with an asymmetrical deepness. In these cases $(\bar{P} \neq 0)$, although the Hamiltonian surface maintains the $p2q2$-symmetry, it becomes inclined by a drif due to the volume constraint $\bar{H}_C^{(V)} = -(\bar{P} \bar{t}^2 /2) q$ (Fig. 4C, 4E and 4G). However, no significant symmetry-breaking is expected from the area constraint $\bar{\xi} \neq 0$, just a shift of the vaccumm energy by $\bar{H}_C^{(V)} = - \bar{\ell}_{\lambda} \bar{\Sigma}$ (Fig. 4A, 4E and 4I).

\section{\label{sec.5} Phase-Space dynamics}

\textit{Hamilton equations.} The time evolution of the membrane-equivalent system is completely defined by two independent equations of motion derived from the Hamiltonian function \cite{21}. The first one tells us about the time-dependence of the generalized velocity $(\dot{q})$: 
\begin{equation}
\label{eq. 18}
\dot{q}(t) = \frac{\partial \bar{H} }{\partial p} = \frac{ (p - p_0) - \bar{\xi}_G q}{\bar{\gamma}}.
\end{equation}

This is equivalent to Eq. $(12)$, previously derived from the Lagrangian in Eq. $(11)$. Because this velocity accounts for the curvature of the cell profile, $\ddot{z} = (R/\tau^2) \dot{q}$, its conjugated moment actually corresponds to the moment of curvature $S = I_1 p$ that describes the local curvatures representing the cell profile (see Section II). For the ideal frictionless case ($\bar{\xi}_G = 0$), one gets $p_{conf} = p - p_0 = \bar{\gamma} \dot{q}$; {\it i.e.} the bare configurational moment exclusively due to curvature $S_{conf} = I_1 p_{conf} = I_1^{(t)} \dot{q} $; here, the time-dependent first moment of inertia is defined as $I_1^{(t)} = \bar{\gamma} (\bar{v}, \bar{t}) I_1$, which stablishes $S_{conf}$ to be a non-conserved quantity. 

More complicated to obtain is the second Hamilton's equation on $\dot{p}$, which involves partial derivatives on the $q$-coordinate. 
\begin{equation}
\label{eq. 19}
\dot{p} = - \frac{\partial \bar{H}}{ \partial q} = - \left( \frac{\partial \bar{H}_{kin}}{ \partial q} + \frac{\partial \bar{H}_{pot}}{ \partial q} + \frac{\partial \bar{H}_{frict}}{ \partial q} + \frac{\partial \bar{H}_C}{ \partial q} \right).
\end{equation} 

This change on generalized momentum corresponds to the net generalized force developed by the system, {\it i.e} $\bar{F} = dp/d\bar{t}$. The total force is the sum of four additive terms $\dot{p} = \bar{F}_{kin} + \bar{F}_{pot} + \bar{F}_{frict} + \bar{F}_C$, which correspond to the four components of the Hamiltonian function in Eq. (13). Although such a force has no dimensions, it stems in a generalized force with the units of a linear momentum; this is $F = \partial H/\partial \dot{z} = (\kappa/\lambda) \partial \bar{H}/\partial q = \mu_0 \lambda \bar{F}$, the impulse imprinted to the membrane-equivalent particle to develop the whole cell profile. The different contributions to this force are defined as follows. 

\textit{Kinetic force.} This is the force put into play to globally accelerate the system. It arises from the $q$-dependent components of the kinetic Hamiltonian, and contains vibrational and translational components:
\begin{equation}
\label{eq. 20}
 \bar{F}_{kin} (p, q, \bar{t}) = \frac{ \partial \bar{H}_{kin}}{\partial q} = - \frac{5p^2 - 6 p p_0 + p_0^2}{2 \bar{\gamma}} \frac{q}{1 + q^2} - \frac{1}{\eta} q,
\end{equation}

The first term describes a highly non-linear force opposed by the dynamic distribution of inertial mass due to vibrational modes, this is $ \bar{F}_{kin}^{(vib)} \sim - p^2 q + o(q^2)$. The second term, $\bar{F}_{kin}^{(trans)} \sim - Mq$ corresponds to the first moment of mass generated upon translation by the material content. Their opposing characters stem on inertia reducing in both cases the net force involved in creating curvature. 

\textit{Potential force.} This is the driving force generated by the curvature-elasticity field. It exclusively arises from the elastic-potential found in the Hamiltonian:
\begin{equation}
\label{eq. 21}
\bar{F}_{pot} (q, \bar{t})= - \frac{ \partial \bar{H}_{pot}}{\partial q} = \bar{K} (q - q_0) \frac{1 - q q_0}{1 + q^2} - \varepsilon_G^2 \bar{K}q \frac{1 + 7q^2/2}{1 + q^2}.
\end{equation}

This genuine potential force is composed by two opposite components; respectively, a positive term due to bending elasticity favouring creation of mean curvature and a negative counterpart due to Gaussian stiffness. These elasticity forces are non-conservative as describe path-dependent stresses. In the particular case of zero spontaneous curvature $(q_0 = 0)$, one gets $ \bar{F}_{pot} \approx \bar{K} (1-\varepsilon_G^2) q + o(q^2)$, which describes the effective linear response of the membrane-equivalent particle in the curvature-elasticity field. For lipid bilayers $(\varepsilon_G \approx 0)$, thus $\bar{F}_{pot} \approx \bar{K} q > 0$ represents the anti-spring force that drives the system to develop the cell profile. The presence of Gaussian elasticity pumping elastic energy from pure flexural modes to shear deformations makes this driving force to effectively decrease as $\bar{K}_{eff} = \bar{K} (1- \varepsilon_G^2)$. At high Gaussian permitivity $(\varepsilon_G^2 >1)$, the potential force becomes indeed a restoring force that opposes to creation of mean curvature, {\it i.e.} $\bar{F}_{pot} < 0$. This points out the determinant role taken by saddle-splay modes in inhibiting curvature deformations due to pure-bending modes. 

\textit{Frictional force.}  This is a non-conservative force opposed to motion, which arises from the frictional component in the Hamiltonian; this is:
 \begin{equation}
 \label{eq. 22}
 \begin{aligned}
 \bar{F}_{frict}  (p, q, \bar{t}) = - \frac{ \partial \bar{H}_{frict}}{\partial q} = - \bar{\xi}_G \frac{(1 + 6 q^2)p + (1 + 4q^2) p_0}{\bar{\gamma}(1 + q^2)}.
\end{aligned}
 \end{equation}

This dissipative force stands on the frictional stresses created upon shear coupling between mean and Gaussian curvatures, $\bar{F}_{frict} \sim - \bar{\xi}_G \dot{q}$; it depends hence on the ammount of curvature momentum $(p_q \sim \dot{q})$. In the linear limit ($|q| << 1$, $\bar{v} \sim 1$), it varies as $ \bar{F}_{frict} \approx - \bar{\xi}_G p/\bar{t} + o(q^2)$, which emphasizes the chief role of frictional stresses in configurating the primogenial history of the membrane ($\bar{F}_{frict} \rightarrow - \infty$ at $\bar{t} \rightarrow 0$). In the propulsion regime $(\bar{\xi}_G < 0)$, this force becomes a driving force able to effectively pump higher curvature $(\bar{F}_{frict} < 0)$.   

\textit{Constraints: Virtual forces.} We finally calculate the last component of the second Hamilton's equation, which describes the virtual forces due to constraints:
 \begin{equation}
 \bar{F}_{C} (q, \bar{t}) = - \frac{\partial \bar{H}_C}{\partial q} = - \frac{\bar{t}}{\bar{v}} \left( \bar{\Sigma} q + \frac{\bar{P} \bar{t} \bar{v}}{2} \right).
 \end{equation}
 
 \begin{figure}
 \includegraphics[width=7.5cm,height=18cm]{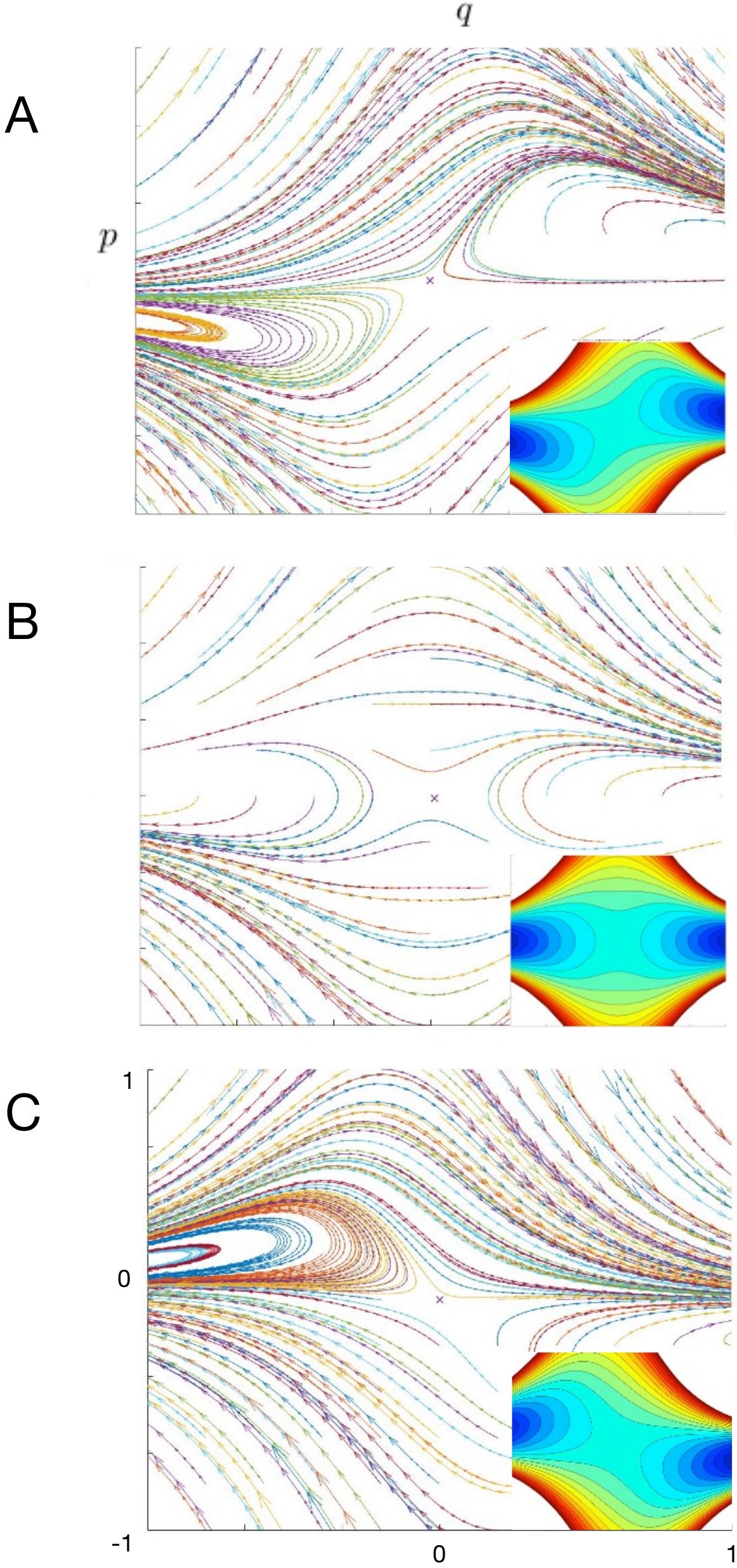}
 \caption[Texto para la lista]{ Phase-space trajectories of the spring-mass membrane equivalent evolving in time (indicated by arrows; the largest the arrow the faster the local velocity of the trajectory). The three cases correspond to different values of the Gaussian susceptibility $\varepsilon_G$, which controls the symmetry of the phase space through frictional drag: A) Positive friction ($\bar{\xi}_G > 0$) at unexpensive creation of Gaussian curvature ($\kappa_G = 0$; $\varepsilon_G = 1$). B) Effectively frictionless ($\bar{\xi}_G =0$); case of typical lipid bilayers with moderate Gaussian rigidity ($\kappa_G = - \kappa$; $\varepsilon_G = 0$). C) Propulsion (or anti-frictional oscillator; $\bar{\xi}_G < 0$); case of high resistence to create saddle points ($\kappa_G = - 2 \kappa$), which propels the system into the inset plots represents level curves of the corresponding energy surfaces completely evolved at the end of the profile pathway (at $\bar{t} = 1$).}
 \label{fig.5}
 \end{figure}
 
These constraint forces are also non-conservative and opposed to motion, as corresponds to virtual forces working to restore the static equilibrium; in the particular case $\bar{F}_C = 0$, the corresponding equation stablishes the equilibrium condition between the virtual forces due to surface tension and cell pressure.

\textit{Phase-space: Dynamic trajectories.} Once the Hamilton's setting is posed as two coupled equations for $\dot{q} (q, p,\bar{t})$ and $\dot{p} (q, p, \bar{t})$, the phase-space $ \{q, p\}$ can be recovered by time-integration. Following a conventional integration schema (Matlab) with the function {\it ode45}, families of phase-space trajectories were computed for different values of initial conditions $(q_{in}, p_{in})$, which determine the total energy involved in the trajectory. As far only harmonic deformations are considered within the CH framework, we will focus on the linear regime on the neighborhood of the metastability pole placed at the configurational center $(p_0, q_0)$. This repulsive center, constitutes the most prominent feature of the phase space, as represents the summit of the repulsive barrier; in phase space, it imposes two orbital classes with an inherent $p2q2$-symmetry. Depending on the ammount of kinetic energy with respect to a critical value $\bar{T}_{crit}$ (which defines the emplacement of a separatrix between two classes of trajectories), we found two dynamical domains: {\it i)} $\bar{H} < \bar{T}_{crit} (p_{in} < p_{crit}, q$), where one founds back-scattered repulsive trajectories confined to the attracting wells; {\it ii)} $\bar{H} >  \bar{T}_{crit}$ ($p > p_{crit}, q$), where open-like trajectories are found asymptotically converging upon effective attraction at high $|q|$. The isoenergetic boundary at $\bar{H}_{crit} (p = p_{crit}, q) = 0$ determines the separatrix between the two orbital regimes. The incline of the separatrix is determined by Gaussian permitivity, which breaks the symmetry of the phase-space in a strength given by the friction coefficient $\bar{\xi}_G$ (see {\it Symmetries}). Notice that the impact of Gaussian friction on the phase-space is formally similar to the damping coefficient in an inverted oscillator. We discuss the particular setting at mass equivalence $(\eta = 1)$, and unconstrained conditions ($\bar{\Sigma} = \bar{P} = \bar{c}_0 = 0$), corresponding to floppy membranes with a natural tendency to adopt the flat configuration. Figure 5 shows three cases specified by the control parameter $\bar{\xi}_G (= \varepsilon_G/\bar{v}^3)$. In every case, the bundles of trajectories retain the same structure as the corresponding Hamiltonian surface (see insets). Particularly, Fig. 5A shows typical phase-space trajectories in the frictional status ($\bar{\xi}_G > 0$), corresponding to membranes prone to create Gaussian curvature ($\varepsilon_G > 0$). In this case, bundles of trajectories remain confined to the assymptotic wells to be later repeled out from the configurational center. Finally, friction dominates again, so the trajectories become further attracted by the opposite well. The higher the positive value of the friction coefficient the more inclined the separatrix, thus the highest the attraction towards the assymptotic limits and the lower the curvatures raised. The ideal frictionless case is shown in Fig. 5B ($\varepsilon_G = 0; \bar{\xi}_G = 0$). Here, the repulsive barrier makes the trajectories to be repeled out from the configurational center to later become assymptotically confined inside the attracting wells; since $\bar{F}_{frict} = 0$ in this case, the $p2q2$-symmetry is not broken. Because frictional forces are absent in this case, the trajectories are now exclusively governed by the repulsive center, which enables the largest curvatures (high $|p|$) at the lowest velocities (low $|q|$). Far away from this center, the trajectories remain confined to the attractive wells (minimizing curvature). Finally, we show in Fig. 5C a typical phase-space in the propulsion regime ($\bar{\xi}_G < 0$), which corresponds to membranes with a high compliance for creating mean curvature at the expense of their resistance to create Gaussian curvature ($\varepsilon_G < -1$). In this case, the self-propeled trajectories are able to escape from the attractive wells and eventually approach the repulsive barrier becoming immediately caught in the assymptotic borderlines at $p \rightarrow p_{0}$. In these cases, the propulsion force makes the escaping trajectories to be stepper with higher strength ($- \bar{\xi}_G$), thus enforcing larger curvatures than in the ideal case.

  \begin{figure}
 \includegraphics[width=8.5cm,height=7cm]{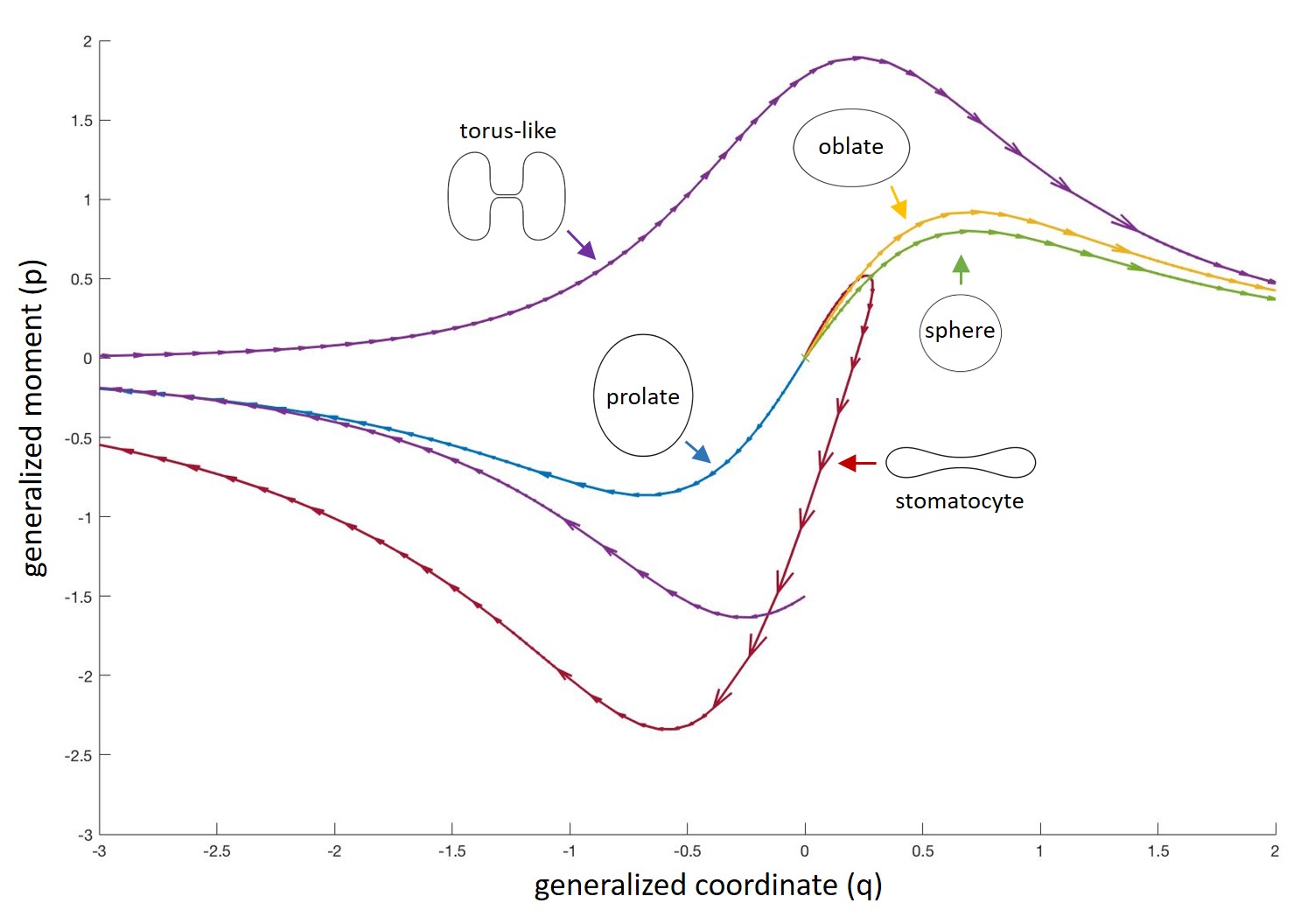}
 \caption[Texto para la lista]{Typical cellular profiles corresponding to phase-space trajectories in the Hamiltonian setting generated from the CH curvature-elasticity field considered in this work (for details, see discussion in Section VI).}
 \label{fig.6}
 \end{figure} 
 
\section{\label{sec.4} MINIMAL ACTION: CELL SHAPES}

The phase-space trajectories above depicted are compatible with the Hamilton's principle of minimal action \cite{21}. For given initial conditions $ (q_{in}, p_{in}) $, they represent the cell profiles (equilibrium) with an instantaneous status given by $p (q_{in}, p_{in}, \bar{t})$ (representing the curvature), and $q (q_{in}, p_{in}, \bar{t})$ (the curve inclination). These phase-space variables are defined in the range of time, $\bar{t} = (0,1]$, corresponding to the definition domain of the radial coordinate, this is $\rho = (0, R]$. At a given instant, the curve characteristics are given as, {\it i)} integrated position $z(\bar{t}) = z(\bar{t} = 0) + R \int_0^{\bar{t}} q (p, \bar{t}')\ d \bar{t}'$, {\it ii)} curve inclination $\dot{z} (\bar{t}) = \lambda q (p, \bar{t})$ and {\it iii)} local curvature $\ddot{z} (\bar{t}) = (\lambda/\tau) \dot{q} (p, \bar{t})$. Such a curve represents the equilibrium cell profile in real space, actually corresponding to the dynamic trajectory of minimal action $(p,q)$ followed by the membrane-equivalent particle in its excursion in phase-space along time. Consequently, for a given set of material parameters $\{ \kappa, \kappa_G, c_0, \Sigma, P \}$, the bundles of parameteric curves $[\rho (\bar{t}), z (\bar{t})]$ compatible with given initial conditions (representing the energy available to the cell). The bending rigidity ($\kappa$) gives the global energy scale, but does not change the structure of the Hamiltonian surface; thus, for a given cell shape, it only determines the cell size. However, the relative contribution of Gaussian elasticity to membrane curvature very much conditions the cell shape, as given by the additional forcing effects included within the permitivity $\varepsilon_G$. The presence of coupling effects $(\varepsilon_G \neq 0)$ induces the mean curvature not only to increase due to an effective softening $\kappa_{eff} = (1- \varepsilon_G^2) \kappa$, but also due to the emergence of frictional dissipation $(\bar{\xi}_G \sim \varepsilon_G)$. Consequently, the structure of the phase space, and the corresponding trajectories, are intrinsically modified by the explicit symmetry breaking introduced by Gaussian friction, or even propulsion (Fig. 5). Other prominent systemic characteristic is the spontaneous curvature, which breaks the symmetry of the Hamiltonian surface, specially when approaching the steady-state (Figs. 4D-F). This parameter induces the trajectories to deviate from the spherical symmetry imprinted by the highly repulsive field during initial inflation.   

Figure 6 shows representative trajectories of the cellular shapes most relevant in the biological setting. Every trajectory only gives the I-quadrant of the cell profile in the $\rho-z$ coordinate system. The complete profile is obtained by reflection along the $\rho$-axis, and the whole cell by rotation around the $z$-axis. For the sake of clarity, we will only discuss the case of unconstrained membranes ($\bar{\Sigma} = \bar{P} = 0$) under initial conditions $q_{in} = p_{in} = 0$ (compatible with zero initial energy, {\it i.e.} $H_{in} = 0$). The sphere is represented by twined trajectories that correspond to orbits of constante curvature; one founds $p = \pm (\bar{t} / \bar{v}^2) (1- q/\bar{v})$ at $\varepsilon_G = 0$, which is compatible with the condition of constant mean curvature (substituting $H = 1/R$ in Eq. $A9$). Spheres can be also found at finite Gaussian permitivity ($\varepsilon_G \neq 0$); in this case, the orbit representing the sphere is compatible with the condition $p = \pm \bar{t}^2/\bar{v} q$ (substituting $K = 1/R^2$ in Eq. $A8$). In both cases, the two solutions ($\pm$) are indeed indiferent, as expected from the metastable character of the configurational center (Fig. 4E). Spheroidal deformation is caused either by introducing systemic constraints ($\bar{\Sigma} \neq 0$, $\bar{P} \neq 0$) or with increasing the absolute value of $c_0$. Precisely, prolate spheroids are found at positive spontaneous curvature ($\bar{c}_0 > 0$), which makes the trajectories to target higher negative moments due to the broken symmetry of the Hamiltonian surface towards the deeper wells in sectors III and IV (Fig. 4D). Conversely, the oblate spheroidal shapes correspond to negative spontaneous curvatures ($\bar{c}_0 < 0$), which makes the energy surface to deform the trajectories to higher positive moments in sectors I and II (Fig. 4F).  A particular case of special biological relevance is the stomacyte-like shapes typical of red blood cells (RBC's), which arise from phase-space trajectories that correspond to highly negative spontaneous curvature. Particularly, for $\bar{c}_0 = -2.4$, standing for the physiological shape of RBCs \cite{22}, a typically flattened discocyte profile (biconvex) is found for initial conditions $p_{in} \approx q_{in} \approx 0$ and given area and volume constraints, respectively $\bar{\Sigma} = 0$ (tensionless) and $\bar{P} = 1000$ (hypertonic). Further, increasing values of the initial energy (compatible with higher initial moment of curvature, e.g. $q_{in} = 0; p_{in} = -1.5)$, give rise to highly deformed stomacytes with unpinched torus-like structure (see Fig. 6). In these cases, the higher energy injected ($H_{in} > 0$) elicits the trajectory to explore complex configurations in phase space, leading to highly deformed cellular shapes.
 
The representative catalogue of cell shapes shown in Fig. 6 allows to envisage the powerfullnes of the Hamiltonian formalism here developed. Given a set of material characteristics, the phase-space obtained for specified initial conditions represents the zoo of cellular shapes that are available for the represented membrane. Our formalism provides indeed a new analytical framework to study the evolution of the equilibrium cellular shapes as a Hamiltonian dynamics in a minimal action phase-space.

\section{\label{sec.7} CONCLUSIONS}

The dynamical phase-space of axisymmetric cells has been reconstructed from a curvature elasticity-field that summarizes the free-energy of the flexible membrane once considered area and volume constraints. This is the usual CH description of minimal cells in terms of membrane elasticity, membrane tension and cell pressure (statics), which has been transformed in an equivalent Hamiltonian setting (dynamics). The geometrical problem has been reconsidered from the point of view of analytical mechanics by reinterpreting the elastic energy of the whole cell as an integrated action arising from a Lagrangian function. The conventional CH free-energy is transformed under invariance conditions, obtaining a minimal Lagrangian described in terms of the generalized characteristics of the curve that represents the cell profile, this is $\bar{L}_{CH} (q, \dot{q}, t)$. The local inclination is identified as the action variable that determines the configurational coordinate $q$, and the local curvature as the configurational velocity $\dot{q}$. Galilean transformation is the key-enabling step that achieves the goal of shifting from the static point of view of an external observer to the inertial frame of reference of a dynamic observer that tracks the membrane profile along a meridian pathway. This transformation renormalizes the radial coordinate into a hidden coordinate that travels together with the meridian observer, thus allowing a drastic reduction of the axysimmetrical problem with redundant degrees of freedom to pure action in generalized coordinates. We additionally explicit the spring-mass interaction by introducing the material content of the membrane and establish a dimensionally-minimal Lagrangian containing kinetic energy and generalized potential. From this minimal Lagrangian built in a one-dimensional basis (the configurational base), the canonical Hamiltonian is constructed in generalized coordinates by taking advantage of the Lagrange transformation. Our construct allows to describe the cellular membrane as an inverted harmonic oscillator driven by bending elasticity and effective friction governed by Gaussian compliance. Conversely, Gaussian stiffness is identified as a propeling force favouring creation of mean curvature. The equations of motion compatible with the Hamilton's principle are finally derived, which has allowed to represent the phase space for given sets of material properties. The trajectories of minimal action are interpreted as the dynamic phase-space description of the cell profiles at equilibrium (minimal free energy, $\delta F = 0$). From this formalism, the phase-space can be used as a global predictor of the cellular shapes in different material settings. The results are equivalent to the classical approach of minimizing the CH free-energy functional. The proposed construction will allow access to much more complex approaches in cell mechanics based on the exploitation of the phase-space dynamics in biological settings.

\section{\label{sec.8} ACKNOWLEDGMENTS}

This research was conducted under grants by Agencia Espa$\tilde{n}$ola de Investigaci$\acute{o}$n (FIS2015-70339-C2-1-R) and Comunidad de Madrid (NANOBIOSOMA-CM; S2013/MIT-2807). AM Maitin acknowledges financial support under grant FIS2015-70339 (AEI). FM is very grateful to Prof. C. Bustamante to host a sabbatical stay at UC Berkeley during the writting of the manuscript, and also to Fulbright Foundation and Ministerio de Educaci$\acute{o}$n, Cultura y Deportes (Spain) to sponsor it as a visiting scholar. The two authors are very grateful to Prof. J.A. Santiago for stimulating discussions.

\appendix

\section{Curvature energy in cylindrical coordinates}

To rationalize the curvature-elasticity model of a minimal cell, we start from Canham-Helfrich's energy in terms of algebraic invariants written as $H=g^{ij} h_{ij}$ and $K=\det h_{ij}/\det g_{ij}$. We consider axial symmetry to describe CH-cells with a revolution profile in the Euclidean space $\mathbb{E}^3$, which will be expressed in terms of the cylindrical coordinates parametrized as:
\begin{equation}
\label{eq.a1}
\begin{cases}
x(\ell ,\phi)=\rho (\ell) \cos \phi \\[0.5ex]
y(\ell, \phi)=\rho (\ell) \sin \phi \\[0.5ex]
z(\ell )=z (\ell). \\[0.5ex]
\end{cases}
\end{equation}
According to differential geometry, this surface is totally defined by two parameters $\ell \in [a,b]$ and $\phi \in (0,2\pi]$. This is the particular parametrization chosen for the revolution surface that represents the membrane of the axysymmetric cell. Such parametrization requires the two surface evolving parameters $ \ell$ and $\phi$; the free-parameter $\ell$ tracks the planar projection of the surface in the $z(\ell)-\rho(\ell)$ plane, whereas the whole cell profile is recovered by revolution of the curve through the angle $\phi$ around the $z$-axis. Under this particular parametrization $ (\ell$, $\phi)$, the axysymmetric cell-shape profile is minimally describable by running the free-parameter $\ell$, which takes the role of a one-dimensional membrane coordinate, or a {\it geometric time}, which describes surface evolution along a meridian. As far the whole surface is assumed with the revolution symmetry, different meridians are expected to be equivalent, both topologically and dynamically. 

In order to obtain the energy as a function of these coordinates, we must use the above definitions for the Gaussian curvature ($K$) and the mean curvature ($H$), that depend on the metric tensor defined by:
\begin{equation}
\label{eq.a2}
g_{ij}=\left< \frac{\partial \vec{r} }{\partial u_i}, \frac{\partial \vec{r} }{\partial u_j} \right>,
\end{equation}
with $i,j=1,2$, being $ \left< \cdot \right> $ the scalar product and $\vec{r}$ a position vector that contains a generic parameterization of the surface in terms of $u_i$. And the tensor of curvature defined by
\begin{equation}
\label{eq.a3}
h_{ij}= \left< \frac{\partial^2 \vec{r} }{\partial u_i \partial u_j} , \hat{N} \right>
\end{equation}
with $\hat{N}$ the normal normalized vector, and the normal is in a point $p$
\begin{equation}
\label{eq.a4}
N(p)=\left. \left( \frac{\partial \vec{r} }{\partial u_1} \times \frac{\partial \vec{r} }{\partial u_2} \right)\right|_{(p)}. 
\end{equation}

Now, we substitute the chosen parameterization $\vec{ r}=[\rho (\ell) \cos \phi,\rho (\ell) \sin \phi,z (\ell)] $, so recalling the symmetry of the metric tensor and where the parametric derivatives $\dot{\rho}=\partial \rho / \partial \ell$, and $\dot{z}=\partial z / \partial \ell$ measure the geometric speed of the curve (inclination) on the direction of the two planar coordinates $( z, \rho)$. Then, the metric tensor read as:
\begin{equation}
\label{eq.a5}
g= 
\begin{pmatrix}
v^2 & 0 \\
0 & \rho^2 
\end{pmatrix}
\end{equation}
with $v=\sqrt{\dot{\rho}^2 + \dot{z}^2}$ being the modulus of the curve velocity, which measures the absolute speed at which the curve evolves in geometric time $\ell$. 

Next, we continue with the normalized vector $\hat{N}=N/||N||$ necessary to calculate the curvature tensor:
 \begin{equation}
 \label{eq.a6}
\hat{N}=\frac{N}{\| N \|}=\frac{(-\dot{z} \cos \phi , -\dot{z} \sin \phi, \dot{\rho}) }{v},
 \end{equation}
which we substitute in Eq. \eqref{eq.a3} and the tensor of curvature reads as:
\begin{equation}
\label{eq.a7}
h= \frac{1}{v}
\begin{pmatrix}
\ddot{z}\dot{\rho}-\ddot{\rho} \dot{z} & 0 \\
0 & \rho \dot{z}
\end{pmatrix}.
\end{equation}

Note that for $v = 0$ this tensor is undefined; this condition represents singular points where the curve speed is nul; by definition, at these points $v = 0$, so any curvature is exactly zero here, this is $h(v=0) \equiv 0$ (flat point). 

With these results at hand, we can get the expressions of the algebraic invariants of the curvature tensor $( H,K)$ in terms of the cylindrical coordinates. Within the parametrization chosen, the expression of Gaussian curvature ($K$) reads as:
\begin{equation}
\label{eq.a8}
K=\frac{\det h}{\det g}=\frac{( \ddot{z}\dot{\rho} - \ddot{\rho}\dot{z})\dot{z}}{v^4\rho}.
\end{equation}

Because the mean curvature is a covariant property of the surface, $H = g^{ij} h_{ij}/2$, we first reverse the metric tensor, {\it i.e.} we must calculate $g^{-1}=g^{ij}$ with Eq. \eqref{eq.a6}. Finally, we obtain the expression for the mean curvature as:
\begin{equation}
\label{eq.a9}
H=g^{ij}h_{ij}=\frac{\ddot{z}\dot{\rho} -\ddot{\rho} \dot{z}}{v^3} + \frac{\dot{z}}{v \rho}.
\end{equation}

Once the algebraic invariants of the curvature tensor have been obtained in terms of cylindrical coordinates (the method is valid for any other coordinates), we can replace the results in Eqs. $A8-A9$ in the formal definition of the Canham-Helfrich curvature-elasticity energy, which reads now in terms of cylindrical coordinates as:
\begin{equation}
\label{eq.a10}
\begin{aligned}
E_c=&\frac{\kappa}{2} \int_{\Omega} d S \ \left(\frac{\ddot{z}\dot{\rho} -\ddot{\rho} \dot{z}}{v^3} + \frac{\dot{z}}{v \rho}-c_0 \right)^2\\
& + \kappa_G \int_{\Omega} dS  \ \frac{( \ddot{z}\dot{\rho} - \ddot{\rho}\dot{z})\dot{z}}{v^4\rho}.
\end{aligned}
\end{equation}

\section{Canham-Helfrich's energy}

The Canham-Helfrich's free-energy considers introduce constant area ($A$) and constant volume ($V$) constraints. To implement them, first we obtain an equation equal to zero for the variation of every constraints, {\it i.e.} $\Delta A = 0$ and $\Delta V = 0$. Then, the variational equations are introduced into the curvature energy with Lagrange multipliers $\Sigma$ and $P$ as $\nabla E_c = \Sigma \nabla A' + P \nabla V'$ and with $\Sigma \Delta A:=\Sigma A'$ and $P\Delta V:=P V'$. To get the explicit expression for the function $E_c$, we integrate the variation obtaining $ E_c= \Sigma  A' + P  V'$. Consequently, the expression for the CH free-energy is:
\begin{equation}
\label{eq.b1}
\begin{aligned}
F_{CH}=&\int_{\Omega} dS\ \left[ \frac{\kappa}{2} \left( H-c_0 \right)^2+ \kappa_G K \right]\\
& + \int_{\Omega} dS\ \Sigma \nabla A'-\int_{\Omega} dS \ P\nabla V'
 \end{aligned}
\end{equation}
which can be rewritten as
\begin{equation}
\label{eq.a20}
F_{CH}=\int_{\Omega} dS\ \left[ \frac{\kappa}{2} \left( H-c_0 \right)^2+ \kappa_G K \right] + \Sigma A'- P V'.
\end{equation}

Under the notation $A' := A$ and $V' := V$, we finally arrive at the Eq. $(1)$ in the main text.

Now, let's consider the constrains in cylindrical coordinates. We started by the area. Let's consider $f$ a generic function that depends on two arbitrary variables denoted by $x_1(\ell)$ and $x_2(\ell)$ with $\ell$ an arbitrary parameter; for this function $f$, the area of the surface obtained by revolution around the $x_2$-axis in parametric form is:
\begin{equation}
\label{eq.b2}
A=2\pi \int_{a}^{b} d \ell \ x_1\sqrt{\dot{x}_1^2+\dot{x}_2^2} .
\end{equation}

Recaling on the symmetry of revolution, we integrate around the angular variable $\phi$; identifying $ x_1 (\ell) $ with $ \rho(\ell) $ and $ x_2 (\ell) $ with $ z(\ell)$, one gets $A = 2 \pi \int d\ell \ \rho v$.

For volume expression, its parametric definition:
\begin{equation}
\label{eq.b3}
V=\pi \int_{a}^{b} d \ell \ x_1^2\ \frac{d x_2}{d \ell}.
\end{equation}

Upon identical assimilation, one gets $V = 2\pi \int d \ell \ \rho^2 \dot{z}/2$. However, the expression for $V$ can be taken in different forms through the Gauss's theorem:
\begin{equation}
\label{eq.b4}
\int_V \nabla \cdotp \mathcal{F} \ dV=\int_S \mathcal{F}\cdotp \hat{n} \ dS
\end{equation}
with $\mathcal{F}$ a vectorial field with constant divergence and $\hat{n}$ normal unit vector defined in Eq. $(A4)$; there, $u_i$ represent the parameters of the surface (matching with $u_1=\phi$ and $u_2=\ell$) and $\vec{r}$ are cylindrical coordinates. The normal vector is $\hat{n}= (\dot{z} \rho \cos \phi, \rho \dot{z} \sin \phi, -\rho \dot{\rho})/(\rho v)$.

If the field has a non-constant divergence, the volume is not be conserved; otherwise, if the divergence of the field is constant, the volume is conserved by correcting the adequate proportionality factor. This geometrical property allows us to choose an arbitrary field with a constant divergence. Let's try with three fields $\mathcal{F}_1= (x,y,z)$, $\mathcal{F}_2 = (x, y, 0)$ and $\mathcal{F}_3 = (0 ,0, z)$, with divergence respectively $3$, $2$ and $1$. Considering these fields in Eq. $(B5)$, we find:
\begin{equation}
\label{eq.b5}
\begin{aligned}
& Field \ \ 1 \ \rightarrow \ 3V = 2 \pi \int d \ell \ \rho (\rho \dot{z} - z \dot{\rho})\\
& Field \ \ 2 \ \rightarrow \ 2V = 2 \pi \int d \ell \ \rho^2 \dot{z} \\ 
& Field \ \  3 \ \rightarrow \ V = 2 \pi \int d \ell \ \rho z \dot{\rho}.
\end{aligned}
\end{equation}

The second equation is chosen to do not include the variable $z$. This geometric property for the integrated volume allows a drastic simplification of the problem.

\section{Canham-Helfrich's action}

Here, we build upon the Canham-Helfrich's action. The surface integral of a scalar field is defined as:
\begin{equation}
\label{eq.c1}
\begin{aligned}
&\int_S dS \ f(x,y,z)= \\
&\int_{\Omega} du dv\ f[x(u,v),y(u,v),z(u,v)]\| \vec{r}_u \times \vec{r}_v\|.
\end{aligned}
\end{equation}
where $\vec{r}_i=\partial \vec{r}/\partial i$ ($i=u,v$) are defined for the position vector $\vec{r}(u,v)=x(u,v)\vec{\imath}+y(u,v)\vec{\jmath}+ z(u,v)\vec{k}$ and $\| \vec{r}_u \times \vec{r}_v\| $ is the normal vector norm $|| n||$. 

In our problem we identify $u$ with $\phi \in (0,2\pi]$, $v$ with $\ell\in [a,b]$ and $\vec{r}(\ell,\phi)=[\rho(\ell)\cos\phi, \rho(\ell)\sin \phi, z(\ell)]$, thus, after substituting in Eq. (A4), we find $|| n ||= \rho v$.

Now, we introduce this result in Eq. \eqref{eq.b1} getting:
\begin{equation}
\label{eq.c2}
\int_S dS \ f(x,y,z)=\int_{\Omega} d\phi d\ell\ f[x(\phi,\ell),y(\phi, \ell),z(\ell)]\rho v.
\end{equation}

Assimilating the left side of this equation with the integral in Eq. \eqref{eq.b1} for the free energy $F_{CH}$ (not depending on the angular variable $\phi$; see Appendix A); consequently, after integration in $\phi$, one gets:
\begin{equation}
\label{eq.c3}
F_{CH}=2\pi \int  d\ell\ f[x(\ell),y(\ell),z(\ell)]\rho v:=2\pi\int d\ell \mathfrak{L}.
\end{equation}

Here, the Lagrangian is identified as $\mathfrak{L}:=f[x(\phi,\ell),y(\phi, \ell),z(\ell)]\rho v$.

\section{Canham-Helfrich's Hamiltonian}

In this section, we will obtain the Hamiltonian from the Legendre transform defined for our case by $\bar{H}(q, p, \bar{t})=\dot{q}p - \bar{L}(q,\dot{q},\bar{t}) = \dot{q} p - \bar{T} + \bar{U}$; with the moment $p$ defined in Eq. $(12)$, we get:
\begin{equation}
\label{eq.d1}
\dot{q} = \bar{\gamma}^{-1} (p - p_0 - \bar{\xi}_G q).
\end{equation}

Thus, the expression of Hamiltonian in terms of $p$ is:
\begin{widetext}
\begin{equation}
\label{eq.d1}
\begin{aligned}
\bar{H}&= p \dot{q} - \bar{T} + \bar{V} + \bar{Q} + \bar{C} = \frac{2 p (p - p_0) - 2 \bar{\xi}_G pq}{2 \bar{\gamma}} - \frac{(p - \bar{\xi}_G q)^2}{2 \bar{\gamma}} - \frac{1 + q^2}{2 \eta} + \bar{V} + \bar{C} - \bar{\xi}_G \frac{(p - p_0)q - \bar{\xi}_G q^2}{\bar{\gamma}} \\
&= \frac{2 p^2 - 2 p p_0}{2 \bar{\gamma}} - \frac{p^2 - 2 \bar{\xi}_G pq + \bar{\xi}_G^2 q^2}{2 \bar{\gamma}} - \frac{2 \bar{\xi}_G pq}{2 \bar{\gamma}} - \frac{1 + q^2}{2 \eta} - \bar{\xi}_G \frac{(p - p_0)q - \bar{\xi}_G q^2}{\bar{\gamma}} + \bar{V} + \bar{C} \\
&= \frac{(p - p_0)^2 - p_0^2}{2 \bar{\gamma}} - \frac{\bar{\xi}_G^2 q^2}{2 \bar{\gamma}} - \frac{1 + q^2}{2 \eta} - \bar{\xi}_G \frac{(p - p_0)q - \bar{\xi}_G q^2}{\bar{\gamma}} + \bar{V} + \bar{C}
\end{aligned}
\end{equation}
\end{widetext}
with $p_0= - \bar{c}_0 \bar{t}/\bar{v}^2$ and $q_0= \bar{c}_0 \bar{t} \bar{v}$.

Then, we define the kinetic term $\bar{H}_{kin}$ as:
\begin{equation}
\label{eq.d2}
\bar{H}_{kin} = \frac{(p - p_0)^2}{2 \bar{\gamma}} - \frac{1 + q^2}{2 \eta}
\end{equation}
and the friction term $\bar{H}_{frict}$ as:
\begin{equation}
\label{eq.d3}
\bar{H}_{frict} = - \bar{\xi}_G \frac{(p - p_0) q}{\bar{\gamma}}.
\end{equation}

Now, we can write the Hamiltonian as:
\begin{equation}
\label{eq.d4}
\begin{aligned}
\bar{H} &= \bar{H}_{kin} + \bar{H}_{frict}- \frac{ p_0^2}{2 \bar{\gamma}} - \frac{\bar{\xi}_G^2 q^2}{2 \bar{\gamma}} + \frac{\bar{\xi}_G^2 q^2}{\bar{\gamma}} + \bar{V} + \bar{C}\\
& = \bar{H}_{kin} + \bar{H}_{frict}- \frac{ p_0^2}{2 \bar{\gamma}} + \frac{\bar{\xi}_G^2 q^2}{2 \bar{\gamma}} + \bar{V} + \bar{C}
\end{aligned}
\end{equation}

Then, we define the potential term $\bar{H}_{pot}$ as:
\begin{equation}
\label{eq.d5}
\begin{aligned}
\bar{H}_{pot} &= \frac{\bar{\xi}_G^2 q^2}{2 \bar{\gamma}} + \bar{V} = \frac{\bar{\xi}_G^2 q^2}{2 \bar{\gamma}} - \frac{\bar{K}}{2} (q - q_0)^2 \\ 
&= \frac{\bar{K}}{2} \varepsilon_G^2 q^2 - \frac{\bar{K}}{2} (q - q_0)^2
\end{aligned}
\end{equation}
and the constraints term $\bar{H}_C$ as:
\begin{equation}
\label{eq.d6}
\begin{aligned}
\bar{H}_C &= - \frac{p_0^2}{2 \gamma} - \bar{C} =  - \frac{p_0^2}{2 \gamma} - \bar{L_0} + \frac{\bar{\ell}_{\lambda} \bar{c}_0^2}{2} =  - \frac{\bar{\ell}_{\lambda} \bar{c}_0^2}{2 }  + \frac{\bar{\ell}_{\lambda} \bar{c}_0^2}{2} - \bar{L_0}\\
& = - \frac{1}{2} \bar{\ell}_{\lambda} \left( 2 \bar{\Sigma} + \bar{P} \frac{\bar{t}}{\bar{v}} q \right).
\end{aligned}
\end{equation}

Finally, the Hamiltonian read as:
\begin{equation}
\label{eq.d7}
\bar{H} = \bar{H}_{kin} + \bar{H}_{pot} + \bar{H}_{frict} + \bar{H}_C
\end{equation}


\nocite{*}

\bibliography{apssamp}

\begin{thebibliography}{99}
\bibitem{62}
No rotational contribution is explicitely considered in this unidimensional problem (as embedded within the angular coordinate $\phi$)
\end{thebibliography}

\end{document}